\documentclass[aps,prd,eqsecnum,showpacs,superscriptaddress,twocolumn,
floatfix,preprintnumbers,altaffilletter]{revtex4}
\usepackage{graphicx,url,amssymb,amsmath,longtable,rotating,color,units,
wasysym,subfigure,wrapfig}

\begin{document}

\title{Long gravitational-wave transients and associated
  detection strategies for a network of terrestrial interferometers}

\author{Eric~Thrane}
\email{ethrane@physics.umn.edu}
\affiliation{School of Physics and Astronomy,
University of Minnesota, Minneapolis, MN 55455, USA}

\author{Shivaraj~Kandhasamy}
\affiliation{School of Physics and Astronomy,
University of Minnesota, Minneapolis, MN 55455, USA}

\author{Christian~D.~Ott}
\affiliation{TAPIR, Caltech, Pasadena, California 91125}

\author{Warren~G.~Anderson}
\affiliation{University of Wisconsin-Milwaukee,
Milwaukee, Wisconsin 53201}

\author{Nelson~L.~Christensen}
\affiliation{Physics and Astronomy, Carleton College,
Northfield, Minnesota 55057}

\author{Michael~W.~Coughlin}
\affiliation{Physics and Astronomy, Carleton College, 
Northfield, Minnesota 55057}

\author{Steven~Dorsher}
\affiliation{School of Physics and Astronomy,
University of Minnesota, Minneapolis, MN 55455, USA}

\author{Stefanos~Giampanis} \affiliation{Albert-Einstein-Institut,
  Max-Planck-Institut f\"ur Gravitationsphysik, D-30167 Hannover,
  Germany.}  \affiliation{University of Wisconsin-Milwaukee,
  Milwaukee, Wisconsin 53201}

\author{Vuk~Mandic}
\affiliation{School of Physics and Astronomy,
University of Minnesota, Minneapolis, MN 55455, USA}

\author{Antonis~Mytidis}
\affiliation{Department of Physics,
University of Florida, Gainsville, FL 32611, USA}

\author{Tanner~Prestegard}
\affiliation{School of Physics and Astronomy,
University of Minnesota, Minneapolis, MN 55455, USA}

\author{Peter~Raffai}
\affiliation{Institute of Physics, E\"otv\"os University, 1117 Budapest, Hungary}

\author{Bernard~Whiting}
\affiliation{Department of Physics,
University of Florida, Gainsville, FL 32611, USA}

\begin{abstract}
  Searches for gravitational waves (GWs) traditionally focus on
  persistent sources (e.g., pulsars or the stochastic background) or on
  transients sources (e.g., compact binary inspirals or 
  core-collapse supernovae), which last for timescales
  of milliseconds to seconds.  
  We explore the possibility 
  of {\it long} GW transients with unknown waveforms
  lasting from many seconds to weeks.
  We propose a novel analysis technique
  to bridge
  the gap between short ${\mathcal O}(\text{s})$ ``burst'' analyses and 
  persistent stochastic 
  analyses.
  Our technique utilizes frequency-time maps of GW strain
  cross-power between two spatially
  separated terrestrial GW detectors.
  The application of our cross-power statistic to searches for GW transients
  is framed as a pattern recognition problem,
  and we discuss several pattern-recognition techniques.
  We demonstrate these techniques by recovering simulated GW
  signals in simulated detector noise.
  We also recover
  environmental noise artifacts, thereby demonstrating a novel technique for
  the identification of such artifacts in GW interferometers.
  We compare the efficiency of this framework to other 
  techniques such as matched filtering.
\end{abstract}

\maketitle

\section{Introduction}
Historically, searches for gravitational-wave (GW) transients fall
into one of two categories: searches for ``bursts'' whose precise
waveforms we cannot predict and searches for compact binary
coalescences, whose waveforms can be predicted (at least for the
inspiral part).  Typically, burst searches focus on events with
$\lesssim \unit[1]{s}$ durations and, indeed, there are many compelling models
for short GW transients (see \cite{all-sky_burst} and references
therein).

In this paper, we put the spotlight on \emph{long} GW transients whose
durations may range from many seconds to weeks. Astrophysical GW
emission scenarios for long transients exist (e.g.,
\cite{ott:09,corsi:09,piro,vanPutten}), 
but their characteristics have not previously been
broadly addressed and no data-analysis strategy 
has been proposed for such events until now.  
(In addition to this work, see recent developments in \cite{tCW}.)
Most of the GW emission
models we consider are burst-like in the sense that the signal
evolution cannot be precisely predicted, however, we refer to
them as ``transients'' to avoid connoting that they are
short-duration.

In Sec.~\ref{astro}, we survey a range of mechanisms for GW emission
that may lead to long transients. These include long-lived
turbulent convection in protoneutron stars (PNSs), rotational instabilities
in rapidly spinning PNSs and in double neutron-star merger remnants, magnetoturbulence
and gravitational instabilities in gamma-ray burst (GRB) accretion torii, 
$r$-modes associated with accreting and newborn neutron stars, 
as well as, perhaps
more speculatively, pulsar glitches and soft-gamma-repeater (SGR) outbursts.

In Sec.~\ref{stats},
we introduce an analysis framework utilizing frequency-time
($ft$)-maps of GW strain cross-power created using data from two or
more spatially separated detectors.  The framework is extended to
include multiple detectors, and we show that it is a generalization of
the GW radiometer algorithm~\cite{radiometer}.  In
Sec.~\ref{gaussianity}, we compare $ft$-cross-power maps of GW data 
(time-shifted to remove astrophysical content)
with Monte Carlo simulations of idealized detector noise.  We shall see that GW
interferometer data is well-behaved enough that thresholds for candidate 
events can be estimated analytically (in at least one case).

In Sec.~\ref{patterns}, we use $ft$-cross-power maps to cast the
search for long GW transients as a pattern recognition problem.  
For the sake of concreteness, we consider two algorithms: a ``box
search''~\cite{box} and a Radon algorithm~\cite{radon_transform}.  In
Sec.~\ref{pem}, we demonstrate the Radon algorithm (as well as
the ``locust'' and Hough algorithms~\cite{raffai}) to identify
environmental noise artifacts in LIGO environmental monitoring
channels---a novel technique for the identification of
such artifacts in GW interferometers.  In
Sec.~\ref{comparison}, we describe how our framework is related to
other detection strategies such as matched filtering.  Concluding
remarks are given in Sec.~\ref{conclusions}.

\section{Astrophysical Sources of Long GW Transients}\label{astro}
In this section, we review a variety of emission mechanisms for long
GW transients.  Most of the mechanisms we consider (summarized in Tab.~\ref{tab:source_summary}) are associated with
one or more of three types of objects: core-collapse supernovae
(CCSNe), compact binary inspirals, or isolated neutron stars.

\begin{table}[tp]
  \begin{tabular}{ll}
    \toprule
    model & source \\
    \toprule
    PNS convection & core collapse \\
    rot. instabilities & BNS coalescence, core collapse, isolated NS \\
    $r$-modes & core collapse, isolated NS \\
    disk instabilities & BNS coalescence, core collapse \\
    high-$\epsilon$ BH binaries & BBH coalescence \\
    pulsar glitches & isolated NS \\
    SGR flares & isolated NS \\
    \toprule
  \end{tabular}
  \caption{Models of long GW transients with associated sources.  BNS and BBH stand for ``binary neutron star'' and ``binary black hole'' respectively. \label{tab:source_summary}}
\end{table}

\subsection{Core-collapse supernovae and long gamma-ray bursts}\label{sec:ccsne}
There is tremendous electromagnetic observational evidence connecting both CCSNe and
long gamma-ray bursts (GRBs) to the core-collapse death of massive stars (see, e.g.,
\cite{wb:06}).  Both are ultimately powered by the release of
gravitational energy, but the precise mechanism by which gravitational
energy is converted into energy of ejecta and radiation is uncertain
in both phenomena (see, e.g., \cite{wb:06,janka:07,ott:09b} and
references therein).  However, all modern models of CCSN and long-GRB
central engines involve violent non-spherical dynamics, making both
systems prodigious emitters of GWs.

The GW signature of CCSNe (recently reviewed in~\cite{ott:09}) may be
composed of contributions from rotating collapse and core bounce
\cite{dimmelmeier:08}, post-bounce protoneutron star (PNS) convection
\cite{dessart:06pns,ott:09,mueller:04}, neutrino-driven convection and
the standing-accretion-shock instability (SASI)
\cite{murphy:09,marek:09b,kotake:09}, PNS pulsations \cite{ott:06prl},
nonaxisymmetric rotational instabilities (both dynamical and secular)
\cite{ott:07prl,ou:04}, asymmetric neutrino emission
\cite{kotake:09,marek:09b,ott:09}, aspherical outflows
\cite{bh:96,fryer:04,shibata:06,takiwaki:10,obergaulinger:06b},
magnetic stresses \cite{takiwaki:10,obergaulinger:06b}, and $r$-mode
pulsations in rotating PNSs (see,
e.g.,~\cite{lindblom:98,bondarescu:09}). Depending on the particular
CCSN mechanism in operation, some emission processes may dominate
while others are suppressed \cite{ott:09b}.

Currently, there are two favored scenarios for the
long-GRB central engine.
In the collapsar scenario~\cite{woosley:93}, 
massive stars collapse to black holes either without an initial
CCSN exploding or via fallback accretion after a successful, but 
weak explosion.
The millisecond-protomagnetar scenario~\cite{bucciantini:09,thompson:04} 
relies on highly magnetized, nascent neutron stars. 
In both cases, any long-GRB activity is
preceded by stellar collapse and a post-bounce phase during which a PNS
exists and GW emission occurs in very similar fashion to regular
CCSNe. In the collapsar scenario, a black hole with an accretion disk
forms. Magnetohydrodynamical processes and/or neutrino pair
annihilation powered by accretion and/or by the extraction of black
hole spin energy eventually launch the GRB jet. GWs may be emitted by
disk turbulence and disk instabilities that may lead to clumping or
disk fragmentation \cite{piro,vanPutten}. In the millisecond-magnetar
scenario, a successful magneto-rotational CCSN explosion (see,
e.g.,~\cite{burrows:07b,dessart:09}) occurs, after which a
high-Lorentz-factor outflow is driven by the millisecond
protomagnetar. GWs may be emitted by convective/meridional currents
and dynamical and secular nonaxisymmetric rotational instabilities in
the protomagnetar~\cite{corsi:09,corsi:09b}.

In CCSNe and in the CCSN-phase of long GRBs most GW emission processes
last until the onset of the CCSN explosion or until PNS-collapse to a
black hole, and hence they have a short duration of order $\lesssim\unit[1-2]{s}$~\cite{fischer:09a,oconnor:10}.  Exceptions are PNS
convection, secular rotational instabilities including $r$-modes and
long-GRB disk/torus instabilities. We discuss these below and
provide order-of-magnitude estimates of their emission characteristics
in the time and frequency domains.

\subsubsection{Protoneutron star convection}
If a CCSN explosion occurs, a stable PNS is left behind and will cool
on a Kelvin-Helmholtz timescale (see, e.g.,~\cite{burrows:86}). Fallback
accretion~\cite{zhang:08}, or, perhaps, a late-time hadron-quark phase
transition (e.g.,~\cite{sumiyoshi:09}) may lead to PNS collapse and
black hole formation. If the PNS survives, a powerful convective
engine, driven by thermal and lepton gradients may continue to operate
for possibly tens of seconds in the cooling phase~\cite{keil:96,dessart:06pns,miralles:00,miralles:04}, making it a long GW transient source.

GW emission estimates from PNS convection are based on results of
simulations that track only the early phase ($\lesssim\unit[1]{s}$ after core bounce)~\cite{mueller:04,ott:09,jm:97}, yet
they have found a number of robust features that translate to later
times. PNS convection occurs at moderate to high Reynolds numbers,
hence, is turbulent and leads to an incoherent, virtually stochastic
GW signal.  Its polarization is random in the nonrotating or slowly
rotating case, but may assume specific polarization due to
axisymmetric rotationally-driven meridional currents in rapidly
spinning PNSs (an effect that remains to be studied in computational
models). In the phase covered by current models, typical GW strains are $h
\sim 3 \times 10^{-23}$ at a galactic distance of
$\unit[10]{kpc}$~\cite{mueller:04,ott:09}.  (``Strain'' refers to the strain 
measured at Earth; strain amplitude
scales like the inverse of the distance from the source.)

While on short timescales, the GW signal of PNS
convection will appear almost as a white-noise burst, its
time-frequency structure is non-trivial, exhibiting a broad
spectral peak at $\mathcal{O}(\unit[100]{Hz})$, which shifts to 
higher frequencies over 
the course of the first second after core
bounce~\cite{mueller:04,ott:09}. This chirp-like trend is likely to
continue for seconds afterward as the PNS becomes more compact. It
should be nearly independent of the size of the convectively unstable
region, since the eddy size will be set by the local pressure scale
height~(see, e.g.,~\cite{thompson:93}).

Based on the $\sim \unit[1.2]{s}$ evolution of a PNS model by
\cite{keil:96,mueller:04}, we expect a total emitted energy of $\sim
1.6\times10^{-10}\,M_\odot c^2$. Assuming that convective GW emission
continues with comparable vigor for tens of seconds, we can scale this to
$E_\mathrm{GW} \sim 4\times 10^{-9} \left( \Delta t/\unit[30]{s}
\right ) \, M_\odot c^2$.

\subsubsection{Rotational instabilities}
Most massive stars ($\gtrsim 98\%$ or so, see~\cite{ott:06spin} and
references therein) are likely to be slow rotators, making PNSs with
birth spin periods of $\sim\unit[10-100]{ms}$.  GRB progenitors, however,
are most likely rapidly spinning, leading to PNSs with birth spins of
$\mathcal{O}(\unit[1]{ms})$ and rotational kinetic energy of up to
$\unit[10^{52}]{erg}$~\cite{burrows:07b}, enough to power a long GRB
through protomagnetar spin-down as suggested by the protomagnetar
model~\cite{bucciantini:09,thompson:04}.  

PNSs, like any self-gravitating (rotating or nonrotating) fluid body,
tend to evolve toward a state of minimal total energy. PNSs are most
likely born with an inner core in solid-body rotation and an outer
region that is strongly differentially rotating~\cite{ott:06spin}.
Magnetorotational instabilities (see, e.g.,~\cite{akiyama:03}) and/or
hydrodynamic shear instabilities (see, e.g.,~\cite{watts:05}) will act
to redistribute angular momentum toward uniform rotation (the
lowest-energy state). The latter type of instability may lead to
significant, though short-term $\tau \lesssim \unit[1]{s}$,
nonaxisymmetric deformation of parts of the PNS and, as a consequence,
to significant GW emission~\cite{ott:07prl,scheidegger:10b,ott:09}.
PNSs in near solid-body rotation that exceed certain values of the
ratio of rotational kinetic to gravitational energy, $T/|W|$, may
deform from an axisymmetric shape to assume more
energetically-favorable triaxial shape of lowest-order $l=m=2$,
corresponding to a spinning bar. Such a bar is a copious emitter of
GWs.  At $T/|W| \gtrsim 0.27$, nonaxisymmetric deformation occurs
dynamically, but will not last longer than a few dynamical times of
$\mathcal{O}(\text{ms})$ (see, e.g.,~\cite{baiotti:07,shibata:05}) due to
rapid angular momentum redistribution, and hence we do not consider its GW
emission in this study.

At $T/|W| \gtrsim 0.14$, a secular gravitational-radiation reaction
or viscosity-driven instability may set in, also leading to
nonaxisymmetric deformation. The timescale for this depends
on the detailed PNS dynamics as well as the details and strength of
the viscosity in the PNS. It is estimated to be $\mathcal{O}(\unit[1]{s})$ for both driving agents, but the expectation is that
gravitational-radiation reaction dominates over
viscosity~\cite{lai:95,lai:01}. The secular instability has the
potential of lasting for $\sim\unit[10-100]{s}$~\cite{lai:95,corsi:09}, and
hence it is of particular interest for our present study.

Once the gravitational-radiation reaction instability sets in, the
initially axisymmetric PNS slowly deforms into $l=m=2$ bar shape and,
in the ideal Dedekind ellipsoid limit, evolves toward zero pattern
speed (angular velocity $\Omega=0$) with its remaining
rotational energy being stored as motion of the fluid in highly
noncircular orbits inside the bar~\cite{lai:95,ou:04}. GW emission
occurs throughout the secular evolution with strain amplitudes $h$
proportional to $\Omega^2$ and to the ellipticity $\epsilon$,
characterizing the magnitude of the bar deformation, leading to an
initial rise of the characteristic strain followed by slow decay as
$\Omega$ decreases~\cite{lai:95,ou:04}.  We expect characteristic
strain amplitudes, defined as $h_c\equiv \tau_\text{GW} f_\text{GW} h$, of
$h_c=\mathcal{O}(\unit[10^{-23} \sim 10^{-22}]{Hz^{-1/2}})$ 
for a source located at $\unit[100]{Mpc}$
and the emission is expected to last at that level for
$\mathcal{O}(\unit[100]{s})$~\cite{corsi:09,lai:95}. The emitted GWs will be
elliptically polarized.

\subsubsection{$R$-modes}
$R$-modes are quasi-toroidal oscillations that have the Coriolis force
as their restoring force. It was shown in~\cite{andersson:98,lindblom:98} 
that $r$-modes in neutron stars are
unstable to growth at all rotation rates by gravitational-radiation
reaction via the secular Chandrasekhar-Friedman-Schutz
instability~\cite{chandrasekhar:70,friedman:78}. $R$-modes emit (at
lowest order) current-quadrupole GWs with $f_\mathrm{GW} = 4/3\,
(\Omega_\mathrm{NS}/2\pi)$ and typical strain amplitudes $h \sim 4.4
\times 10^{-24} \alpha (\Omega_\mathrm{NS} / \sqrt{\pi G \bar{\rho}})^3
(\unit[20]{Mpc}/D)$~\cite{owen:98}, where $\Omega_\mathrm{NS}$ is
the NS angular velocity, $D$ is the distance to the source and
$\bar{\rho}$ is the mean neutron star density. The parameter $\alpha \in [0,1]$
is the dimensionless saturation amplitude of the $r$-modes and its
true value has been the topic of much debate. Most recent work
suggests (see~\cite{bondarescu:09,bondarescu:07} and references
therein) that $\alpha \ll 0.1$ and, perhaps, does not exceed $\sim
10^{-5}$ due to non-linear mode coupling effects~\cite{brink:04}.
Generally, $r$-modes are expected to be a source of very-long-lasting
quasi-continuous GW emission, though
long GW transients may be possible in the
case of high saturation amplitudes (e.g.,~\cite{owen_rmodes}).

Potential astrophysical sources of GWs from $r$-modes are accreting
neutron stars in low-mass X-ray binaries (e.g.,
\cite{lars,lmxb,bondarescu:07}) and, more relevant in the present
context, newborn, rapidly spinning neutron
stars~\cite{owen:98,lindblom:98,bondarescu:09}. In the latter,
$r$-modes may play an important role in the early spin
evolution~\cite{bondarescu:09,crab_rmodes}.

\subsubsection{Accretion disk instabilities}
In the long GRB collapsar scenario, the central engine consists
of a black hole surrounded by an accretion
disk/torus~\cite{woosley:93,wb:06}. The inner part of the disk is
likely to be sufficiently hot to be neutrino cooled and
thin~\cite{popham:99} while the outer regions with radius $r\gtrsim\,
50 R_\mathrm{S} = 100\, G M_\mathrm{BH} c^{-2}$ are cooled
inefficiently and form a thick accretion torus \cite{popham:99,piro}.
A variety of (magneto)-hydrodynamic instabilities may occur in the
disk/torus leading to the emission of GWs.

Piro and Pfahl~\cite{piro} considered gravitational instability of the
outer torus leading to fragmentation facilitated by efficient cooling
through helium photodisintegration. Multiple fragments may collapse to
a single big dense fragment of up to $\sim 1 M_\odot$ that then
travels inward either by means of effective viscosity and/or GW
emission. In both cases, the inspiral will last 
$\mathcal{O}(\unit[10-100]{s})$, making 
it a prime candidate source for a long GW
transient. Piro and Pfahl predict maximum dimensionless strain
amplitudes $|h| \sim 2 \times 10^{-23} (f_\text{GW}/\unit[1000]{Hz})^{2/3}$ for
emission of a system with a fragment mass of $1 M_\odot$, a black hole
mass of $8 M_\odot$, and a source distance of $\unit[100]{Mpc}$. The
frequency will slowly increase 
over the
emission interval, making the emission quasi-periodic and, thus,
increasing its detectability by increasing its characteristic strain
$h_c $ up to $\mathcal{O}(\times 10^{-22})$ at $f_\text{GW} \sim
\unit[100]{Hz}$ and a source distance of $\unit[100]{Mpc}$.

Van~Putten, in a series of articles~(see,
e.g.,~\cite{vanputten:01,vanPutten,vanputten:08} and references
therein), has proposed an extreme ``suspended--accretion'' scenario in
which the central black hole and the accretion torus are dynamically
linked by strong magnetic fields. In this picture, black hole
spin-down drives both the actual GRB central engine and strong
magnetoturbulence in the torus, leading to a time-varying mass
quadrupole moment and, thus, to the emission of GWs.  Van~Putten
postulated, based on a simple energy argument, branching ratios of
emitted GW energy to electromagnetic energy of
$E_\mathrm{GW}/E_\mathrm{EM} \gtrsim 100$ and thus, $E_\mathrm{GW} \sim
\unit[\mathrm{few}\times 10^{53}]{erg}$.  These numbers are perhaps
unlikely to obtain in nature, but the overall concept of
driven magnetoturbulence is worth considering.

In van Putten's theory, the magnetoturbulent torus excitations produce
narrowband elliptically polarized GWs with a frequency between
$(\unit[1\sim2]{kHz})(1+z)$ for a GRB at a redshift of
$z$~\cite{vanputten:01}.  The frequency is predicted to vary with time
such that $df/dt = \mathrm{const}$~\cite{vanputten:01}.  GW emission
continues for approximately the same duration as the electromagnetic
emission, lasting typically for seconds to minutes \cite{wb:06}.  
With van Putten's energetics, a long GRB at a distance of
$\unit[100]{Mpc}$ is predicted to produce a strain of
$h\sim10^{-23}$, which is comparable to the expected Advanced LIGO
noise at $\unit[1000]{Hz}$~\cite{vanputten:01}.  Integrating many
seconds of data, such a loud signal should stand out above the
Advanced LIGO noise, making it likely that a strong statement can be
made about this model in the advanced detector era.

\subsection{Postmerger evolution of double neutron-star coalescence}
In Subsec.~\ref{sec:ccsne}, we discussed a variety of
scenarios for long GW transients in the context of PNS and black-hole
-- accretion-disk systems left in the wake of CCSNe and in collapsars.
A similar situation is likely to arise in the postmerger stage of
double neutron-star coalescence. The initial remnant will be a hot supermassive
neutron star that, depending on the mass of the binary constituents and
on the stiffness of the nuclear equation of state, may
survive for hundreds of milliseconds (e.g., \cite{kiuchi:09} and
references therein). In these systems, many of the GW emission
mechanisms discussed in the stellar collapse scenario may be active.
Hence, it may be fruitful to search for long GW transients following
observed inspiral events as well as following short GRBs, (which are
expected to be associated with binary inspirals).

Inspiral events, however, need not invoke the formation of a PNS in
order to produce a long GW transient.  Highly eccentric black hole
binary inspirals are expected to produce complicated waveforms that
are difficult to model with matched filtering and may persist for
hundreds of seconds~\cite{birjoo,janna}, and thus they are suitable
candidates for long GW transient searches.
According to some models~\cite{BBH_in_GN}, a significant fraction of BBH form dynamically with high eccentricities ($\epsilon>0.9$) leading to an Advanced LIGO event rate of $\sim\unit[1-100]{yr^{-1}}$.
Given such high rates, it is highly likely that these models can be thoroughly probed in the advanced detector era.

\subsection{Isolated neutron stars}
Isolated neutron stars are another potential source of long GW
transients.  In the following, we discuss pulsar glitches and 
soft-gamma repeater flares as potential sources of GWs.

Pulsar glitches are sudden speed-ups in the rotation of pulsing
neutron stars observed by radio and X-ray observatories.  The
fractional change in rotational frequency ranges from $10^{-10}<\Delta
f/f<5\times10^{-6}$, corresponding to rotational energy changes of
$\lesssim\unit[10^{43}]{erg}$~\cite{14pulsars,pglitch_scaling}.  The
speed-up, which takes place in $<\unit[2]{min}$, is followed by a
period of relaxation (typically weeks) during which the pulsar slows
to its pre-glitch frequency~\cite{pglitch_nature}.

The mechanism by which pulsar glitches occur is a matter of ongoing
research~\cite{baym,anderson_glitch,ruderman,takatsuka,mochizuki}, and
the extent to which they emit GWs is unknown.  We therefore follow
Andersson et al.~\cite{pglitch_strain} and assume that the emitted
energy in GWs is comparable to the change in rotational energy.  Given
these energetics, and assuming a simple exponentially-decaying damped
waveform, a nearby ($d=\unit[1]{kpc}$) glitch can produce, e.g., a
$\mathcal{O}(\unit[10]{s})$ quadrupole excitation with a strain of
$h\sim8\times10^{-24}$ at $\unit[3.8]{kHz}$~\cite{pglitch_strain}.
This is about six times below the Advanced LIGO noise floor, which
effectively rules out the possibility of detection.  A long
GW event measured by Advanced LIGO and coincident with a pulsar glitch
would therefore suggest a radically different glitch mechanism than 
the one considered in~\cite{pglitch_strain}.

Flares from soft-gamma repeaters (SGRs) and anomalous X-ray pulsars
(AXPs), which may be caused by seismic events in the crusts of
 magnetars, have
also been proposed as sources of GWs.  Recent searches by LIGO have
set limits on lowest-order quadrupole ringdowns in SGR
storms~\cite{ligo_sgr} and in single-SGR events \cite{ligo_sgr:08}.
SGR giant flares are associated with huge amounts of electromagnetic
energy ($\unit[10^{44}-10^{46}]{erg}$), and they are followed by a
$\mathcal{O}(\unit[100]{s})$-long tail characterized by quasi-periodic
oscillations (see, e.g.,~\cite{glitch_facts}).  It is hypothesized
that quasi-periodic oscillations following SGR flares may emit GWs
through the excitation of torsional modes \cite{ligo_sgr_qpo:07}.

Current models of GW from 
SGRs/AXPs~\cite{glampedakis,samuelsson,levin,sotani,owen_magnetar,horvath,deFreitas,ioka} are very
preliminary, but even if we assume that only $0.1\%$ of the
$\unit[10^{46}]{erg}$ of electromagnetic energy in a nearby 
SGR flare is converted into GWs, then SGRs are
nonetheless attractive targets in the advanced detector era.  Current
experiments have set limits on the emission of GW energy from
SGRs (in the form of short bursts) at a level of
$E_\text{GW}\lesssim\unit[10^{45}]{erg}$~\cite{ligo_sgr,ligo_sgr:08}
($\approx10\%$ of the electromagnetic energy of a giant flare
depending on the waveform
type).  Since sensitivity to $E_\text{GW}\propto h^2$, it is likely
that we can probe interesting energy scales ($E_\text{GW}\approx0.1\%
\, E_\text{EM}$) in the advanced detector era.

\section{An excess cross-power statistic}\label{stats}
\subsection{Definitions and conventions}
Our present goal is to develop a statistic $\hat{\cal Y}_\Gamma$ which can be used to estimate the GW power $H_\Gamma$ (or the related quantities of GW fluence $F_\Gamma$ and energy $E_\Gamma$) associated with a long GW transient event confined to some set of discrete frequencies and times $\Gamma$.
In order to define GW power, we first note the general form of a point-source GW field in the transverse-traceless gauge:
\begin{equation}\label{eq:signal_model}
  \begin{split}
    h_{ab}(t,\vec x) = \sum_A \int_{-\infty}^\infty df \,
    e^A_{ab}(\hat\Omega) \tilde h_A(f)
    \, e^{2\pi i f(t+\hat\Omega\cdot\vec x /c)} .
  \end{split}
\end{equation}
Here $\hat\Omega$ is the direction to the source, $A$ is the polarization state and $\{e_{ab}^A\}$ are the GW polarization tensors with Cartesian indices $ab$, (see App.~\ref{GW_power} for additional details).
Since Eq.~\ref{eq:signal_model} describes an astrophysical source, the Fourier transform of the strain $\tilde h(f)$ is defined in the continuum limit.

We now, however, consider a discrete measurement on the interval between $t$ and $t+T$ measured with a sampling frequency of $f_s$, which corresponds to $N_s=f_s T$ independent measurements.
We utilize a discrete Fourier transformations denoted with tildes:
\begin{equation}\label{eq:DFT}
  \begin{split}
    \tilde q_k \equiv & \frac{1}{N_s} \sum_{n=0}^{N_s-1} q_n e^{-2\pi ink/N_s} \\
    q_n \equiv & \sum_{k=0}^{N_s-1} \tilde q_k e^{2\pi ink/N_s} .
  \end{split}
\end{equation}
The variable $t$---separated from other arguments with a semicolon---refers to the segment start time, as opposed to individual sampling times, denoted by $t$ with no semicolon, e.g., $s(t)$.
Variables associated with discrete measurements are summarized in Tab.~\ref{tab:DFT}.

\begin{table}[tp]
  \begin{tabular}{ll}
    \toprule \\
    $f_s$ & sampling frequency \\
    $\delta t \equiv 1/f_s$ & sampling time \\
    $t$ & segment start time \\
    $T$ & segment duration \\
    $\delta f$ & frequency resolution \\
    $N_s$ & number of sampling points in one segment \\
    \toprule
  \end{tabular}
  \caption{Variables describing discrete measurements. \label{tab:DFT}}
\end{table}

The GW strain power spectrum (measured between $t$ and $t+T$ with a sampling frequency $f_s$ in a frequency band between $f$ and $f+\delta f$) is:
\begin{equation}\label{eq:spectrum}
  H_{AA'}(t;f) = 
  2
  \langle \tilde{h}_A^*(t;f) \tilde{h}_{A'}(t;f) \rangle .
\end{equation}
The factor of $2$ comes from the fact that $H_{AA'}(t;f)$ is the one-sided power spectrum.

It is convenient to characterize the source with a single spectrum that includes contributions from both $+$ and $\times$ polarizations.
We therefore define
\begin{equation}\label{eq:power}
  H(t;f) \equiv \text{Tr}\left[H_{AA'}(t;f)\right] ,
\end{equation}
so as to be invariant under change of polarization bases.
This definition is a generalization of the one-sided power spectrum for unpolarized sources found in~\cite{allen-romano,radiometer,sph}.

Our estimator $\hat {\cal Y}_\Gamma(\Omega)$ utilizes frequency-time ($ft$)-maps: arrays of pixels each with a duration determined by the length of a data segment $T$ and by the frequency resolution $\delta f$.
In Subsec.~\ref{multi}, we describe how $\hat{\cal Y}_\Gamma$ can be constructed by combining {\em clusters} of $ft$-map pixels.
We thereby extend the stochastic-search formalism developed in~\cite{allen-romano,radiometer,sph} beyond models of persistent unpolarized sources to include polarized and unpolarized transient sources.
In doing so, we endeavor to bridge the gap between searches for short ${\cal O}(\mathrm{s})$ signals and stochastic searches for persistent GWs.
We begin by considering just one pixel in the $ft$-map.

\subsection{A single $ft$-map pixel}\label{single_pixel}
In App.~\ref{Y_derivation}, we derive the form of an estimator $\hat Y$ for GW power $H(t;f)$ in a single $ft$-pixel by cross-correlating the strain time series $s_I(t)$ and $s_J(t)$ from two spatially separated detectors, $I$ and $J$, for a source at a sky position $\hat\Omega$~\footnote{For the sake of simplicity, the calculations in this section ignore effects from windowing and the use of overlapping segments.}.
We find that
\begin{equation}\label{eq:pixelY}
  \hat Y(t;f,\hat{\Omega}) \equiv  
  \text{Re}\left[
    \tilde{Q}_{IJ}(t;f,\hat\Omega) \,
    C_{IJ}(t;f)
    \right] .
\end{equation}
Here, $C_{IJ}(t;f)$ is the one-sided cross-power spectrum
\begin{equation}
  C_{IJ}(t;f)\equiv 2 \, \tilde{s}_I^*(t;f) \tilde{s}_J(t;f) .
\end{equation}
Meanwhile, $\tilde{Q}_{IJ}(t;f,\hat\Omega)$ is a filter function, which depends, among other things, on the source direction and polarization.
For unpolarized sources (see App.~\ref{Y_derivation}),
\begin{equation}
  \tilde Q_{IJ}(t;f,\hat\Omega) = \frac{1}{\epsilon_{IJ}(t;\hat\Omega)} 
  e^{2 \pi i f \hat{\Omega} \cdot \Delta \vec{x}_{IJ} /c} .
\end{equation}
where $\epsilon_{IJ}(t;\hat\Omega)\in[0,1]$, the ``pair efficiency,'' is
\begin{equation}\label{eq:efficiency}
  \epsilon_{IJ}(t;\hat\Omega) \equiv \frac{1}{2} \sum_A F_I^A(t; \hat\Omega) 
  F_J^{A}(t; \hat\Omega) .
\end{equation}
Here $F_I^A(t; \hat\Omega)$ is the ``antenna factor'' for detector $I$ and $\Delta \vec x_{IJ}\equiv \vec x_I - \vec x_J$ is the difference in position vectors of detectors $I$ and $J$; (see App.~\ref{GW_power}).
Pair efficiency is defined such that a GW with power $H$ will induce $IJ$ strain cross-power given by $\epsilon_{IJ} H$.
It is unity only in the case where both interferometers are optimally oriented so that the change in arm length is equal to the strain amplitude.
For additional details (including a derivation of the pair efficiency for polarized sources) see Apps.~\ref{GW_power}, \ref{Y_derivation} and~\ref{polarized}.

The variance of $\hat Y$ is calculated in App.~\ref{variance}.
Then in App.~\ref{sigma_hat}, we show that the following expression for $\hat\sigma_Y^2(t;f,\hat\Omega)$ (motivated by analogy with stochastic analyses~\cite{allen-romano}) is an estimator for the variance of $\hat Y$,
\begin{equation}\label{eq:pixelvar}
  \begin{split}
    \hat\sigma_Y^2(t;f,\hat\Omega) = \frac{1}{2}
    |\tilde Q_{IJ}(t;f,\hat\Omega)|^2 
    P_I^\text{adj}(t;f)P_J^\text{adj}(t;f) ,
  \end{split}
\end{equation}
where $P_I^\text{adj}$ is the average one-sided auto-power spectrum in neighboring pixels,
\begin{equation}\label{eq:PSD_def}
  P_I^\text{adj}(t;f) \equiv 2 \,
  \overline{ \left| \tilde s_I(t;f) \right|^2 } .
\end{equation}
The overline denotes an average over neighboring pixels~\footnote{In order to chose a suitable number of neighboring pixels to average over, one must typically take into account the stationarity of the detector noise.  This discussion, however, is outside our current scope.}.

From Eqs.~\ref{eq:pixelvar} and \ref{eq:pixelY}, we define the signal to noise ratio $\text{SNR}(t;f,\hat\Omega)$ for a single $ft$-map pixel:
\begin{equation}\label{eq:pixelSNR}
  \begin{split}
    \text{SNR}(t;f,\hat\Omega) \equiv & \, \hat Y(t;f,\hat\Omega)/
    \hat\sigma_Y(t;f,\hat\Omega) \\
    = & 
    \text{Re} \left[
      \frac{\tilde Q_{IJ}(t;f,\hat\Omega)}
	   {\left|\tilde Q_{IJ}(t;f,\hat\Omega)\right|}
	   \frac{C_{IJ}(t;f)}
		{\sqrt{\frac{1}{2} P_I^\text{adj} P_J^\text{adj}}}
		\right]
  \end{split}
\end{equation}
It depends on the phase of $\tilde Q_{IJ}(t;f,\hat\Omega)$, but not on the magnitude.
Thus, a single $ft$-pixel taken by itself contains no information about the polarization properties of the source, since the polarization does not affect the phase of $\tilde Q$.
This degeneracy is broken when we combine $ft$-pixels from different times or from different detector pairs.

\subsection{Energy, fluence and power}
One of the most interesting intrinsic properties of a transient source of GWs is the total energy emitted in gravitational radiation, $E_\text{GW}$.
By measuring $E_\text{GW}$ (and, when possible, comparing it to the observed electromagnetic energy, $E_\text{EM}$), we can make and test hypotheses about the total energy associated with the event as well as constrain models of GW production.
Thus, it is useful to relate $\hat Y(t;f,\hat\Omega)$ to $E_\text{GW}$ and the related quantity of fluence.
If the GW energy is emitted isotropically (in general it is not) then~\cite{ShapiroTeukolsky},
\begin{equation}
  E_\text{GW} = 4\pi R^2 \frac{c^3}{16\pi G} \int dt \left( 
  \dot h_+^2(t) + \dot h_\times^2(t)
  \right) ,
\end{equation}
where $R$ is the distance to the source.
It follows that the equivalent isotropic energy is related to our cross-power estimator as follows:
\begin{equation}
  \hat E_\text{GW}(t;f,\hat\Omega) = 4\pi R^2 \frac{\pi c^3}{4G} 
  \left(T f^2 \right)
  \hat Y(t;f,\hat\Omega) .
\end{equation}

$\hat E_\text{GW}(t;f,\hat\Omega)$ may contain significant uncertainty about the distance to the source or the isotropy of the GW emission.
It is therefore useful to define a statistic that contains only uncertainty associated with the strain measurement.
The natural solution is to construct a statistic for GW fluence, $\hat F_\text{GW}(t;f,\hat\Omega)$, which is given by
\begin{equation}
  \begin{split}
    \hat F_\text{GW}(t;f,\hat\Omega) = & 
    \frac{\hat E_\text{GW}(t;f\hat\Omega)}{4\pi R^2} \\
    = & T f^2 \left(\frac{\pi c^3}{4G}\right)
    \hat Y(t;f,\hat\Omega) .
  \end{split}
\end{equation}
In the subsequent section, we show how multiple pixels can be combined to calculate the average power inside some set of pixels.
The same calculation can be straightforwardly extended to calculate the total fluence.
This is done by reweighting $\hat Y(t;f,\hat\Omega)$ and $\hat \sigma(t;f,\hat\Omega)$ by $(\pi c^3/4G)(T f^2)$.
Also Eqs.~\ref{eq:GammaY} and~\ref{eq:GammaVar} must be scaled by the number of pixels in a set, $N$; (otherwise we obtain average fluence instead of total fluence).

\subsection{Multi-pixel statistic}\label{multi}
We now generalize from our single-pixel statistic to accommodate transients persisting over $N$ pixels in some set of pixels, $\Gamma$.
We define $H_\Gamma$ to be the average power inside $\Gamma$,
\begin{equation}\label{eq:model_Gamma}
  H_\Gamma \equiv \frac{1}{N} \sum_{t;f\in\Gamma} H(t;f) .
\end{equation}
A minimum-variance estimator for the GW power in $\Gamma$ can be straightforwardly constructed from a weighted sum of $\hat Y(t;f,\hat\Omega)$ for each pixel in $\Gamma$,
\begin{equation}\label{eq:GammaY}
  \hat{\cal Y}_\Gamma(\hat\Omega) =
  \frac{\sum_{t;f\in\Gamma} \hat Y(t;f,\hat\Omega) \,
    \hat\sigma_Y(t;f,\hat\Omega)^{-2}}
  {\sum_{t;f\in\Gamma}\hat\sigma_Y(t;f,\hat\Omega)^{-2}} .
\end{equation}
Here we assume that the power is either evenly or randomly distributed inside $\Gamma$, which is to say $\langle H(t;f) \rangle = \langle H(t';f') \rangle \equiv H_0$ and so $\langle H_\Gamma \rangle = H_0$.
Thus, 
\begin{equation}
  \begin{split}
    \langle \hat{\cal Y}_\Gamma(\hat\Omega) \rangle & =
    \left\langle
    \frac{\sum_{t;f\in\Gamma} \hat Y(t;f,\hat\Omega) 
      \, \hat\sigma_Y(t;f,\hat\Omega)^{-2}}
	 {\sum_{t;f\in\Gamma}\hat\sigma_Y(t;f,\hat\Omega)^{-2}} 
	 \right\rangle \\
	 & =
	 \frac{\sum_{t;f\in\Gamma} \langle \hat Y(t;f,\hat\Omega) \rangle
	   \, \hat\sigma_Y(t;f,\hat\Omega)^{-2}}
	 {\sum_{t;f\in\Gamma}\hat\sigma_Y(t;f,\hat\Omega)^{-2}} \\
	 & =
	 H_0 
	 \left( \frac{\sum_{t;f\in\Gamma} \hat\sigma_Y(t;f,\hat\Omega)^{-2}}
	 {\sum_{t;f\in\Gamma}\hat\sigma_Y(t;f,\hat\Omega)^{-2}} \right)
	 = \langle H_\Gamma \rangle .
  \end{split}
\end{equation}

Here we have additionally assumed that there are no correlations between $\hat Y(t;f,\hat\Omega)$ in different pixels.
If the GW signal in different pixels is correlated, then the $\{\hat Y(t;f,\hat\Omega)\}$ are correlated and Eq.~\ref{eq:GammaY} should, in theory, be modified to include covariances between different pixels.
In practice, however, the covariance matrix is not known, and so we must settle for this approximation, which gives the estimator a higher variance than could be achieved if the covariance matrix was known.

The associated estimator for the uncertainty is
\begin{equation}\label{eq:GammaVar}
  \hat \sigma_\Gamma(\hat\Omega)=
  \left(\sum_{t;f\in\Gamma}\hat\sigma_Y(t;f,\hat\Omega)^{-2}\right)^{-1/2} .
\end{equation}
The choice of the set of pixels $\Gamma$ to include in the sum in Eq.~\ref{eq:GammaY} is determined by the signal model.
For example, a slowly varying narrowband signal can be modeled as a line of pixels on the $ft$-map.
We explore this and other choices for $\Gamma$ in greater detail in Sec.~\ref{patterns}.

The $\text{SNR}$ for given a set of pixels $\Gamma$ is given by
\begin{equation}\label{eq:GammaSNR}
  \text{SNR}_\Gamma(\hat\Omega) = \frac{\hat{\cal Y}_\Gamma(\hat\Omega)}{\hat\sigma_\Gamma(\hat\Omega)} .
\end{equation}
Since $\text{SNR}_\Gamma$ is the weighted sum of many independent measurements, we expect, due to the central limit theorem, that the distribution of $\text{SNR}_\Gamma$ will be increasingly well-approximated by a normal distribution as the volume of $\Gamma$ increases and more pixels are included in the sum~\footnote{Here we also assume that the probability density function for each pixel is the same, which is to say that the noise and signal are approximately stationary.}.

\subsection{Multi-detector statistic}
It is straightforward to generalize $\hat{\cal Y}_\Gamma$ for a detector network ${\cal N}$ consisting of $n\geq2$ spatially separated detectors.
First, we generate $n(n-1)/2$ $ft$-maps for each pair of interferometers.
Then we extend the sum over pixels in Eq.~\ref{eq:GammaY} to include a sum over unique detector pairs $p(I,J)$:
\begin{equation}\label{multi-det_Y}
  \hat{\cal Y}^{\cal N}_\Gamma(\hat\Omega)=\frac{\sum_{p(I,J)}
    \sum_{t;f\in\Gamma} \hat Y_{IJ}(t;f,\hat\Omega) 
    \hat\sigma_{IJ}(t;f,\hat\Omega)^{-2}}
  {\sum_{p(I,J)}\sum_{t;f\in\Gamma}\hat\sigma_{IJ}(t;f,\hat\Omega)^{-2}} .
\end{equation}
By construction, the expectation value is
\begin{equation}
  \langle \hat{\cal Y}_\Gamma^{\cal N} \rangle = H_\Gamma .
\end{equation}
The associated uncertainty is
\begin{equation}\label{eq:multi-det_sigma}
  \hat\sigma^{\cal N}_{\cal Y}(\hat\Omega)=
  \left(\sum_{p(I,J)}\sum_{ft}
  \hat\sigma_{IJ}(t;f,\hat\Omega)^{-2}\right)^{-1/2} .
\end{equation}

Adding new detectors to the network improves the statistic by mitigating degeneracies in sky direction and polarization parameters and also by improving sensitivity to $H_\Gamma$ by increasing the number of pixels contributing to $\hat{\cal Y}_\Gamma^{\cal N}$.

\subsection{Relationship to the GW radiometer}
The multi-pixel statistic $\hat{\cal Y}_\Gamma$ is straightforwardly related to the GW radiometer technique, which has been used to look for GWs from neutron stars in low-mass X-ray binaries~\cite{radiometer}.
By constructing a rectangular set of pixels consisting of one or more frequency bins and lasting the entire duration of a science run, we recover the radiometer statistic as a special case.

It is instructive to compare the unpolarized radiometer statistic~\cite{radiometer} with our $\hat{\cal Y}_\Gamma$:
\begin{equation}
  \begin{split}
    \hat Y^\text{rad}(t;f,\hat\Omega) \equiv \int_{-\infty}^\infty df \tilde 
    Q_{IJ}^\text{rad}(t;f,\hat\Omega) \\
    \tilde s_I^\star(t;f) \tilde s_J(t;f)
  \end{split}
\end{equation}
\begin{eqnarray}
  \tilde Q^\text{rad}_{IJ}(t;f,\hat\Omega)  & \equiv & 
  \lambda_t \frac{\gamma(t;f,\hat\Omega) \bar H(f)}{P_I(f) P_J(f)} \\
  \gamma_{IJ}(t;f,\hat\Omega) & \equiv & 
  \epsilon_{IJ}(t;\hat\Omega) \,
  e^{2\pi i f \hat\Omega\cdot\Delta \vec{x}_{IJ}/c} .
\end{eqnarray}
Here $\gamma_{IJ}(t;f,\hat\Omega)$ is the so-called overlap reduction factor, $\lambda_t$ is a normalization factor and $\epsilon_{IJ}(t;\hat\Omega)$ is the pair efficiency, which we define in Eq.~\ref{app_eq:efficiency} and Eq.~\ref{eq:efficiency}.

There are two things worth noting here.
First, the extra factor of $H(f)/P_I(f)P_J(f)$ in the expression for $\tilde Q^\text{rad}_{IJ}$ does not appear in our expression for $\tilde Q_{IJ}$ (see Eq.~\ref{app_eq:Q_unpolarized}).
The factor of $1/P_I(f)P_J(f)$ is proportional to $\sigma(f)^{-2}$, and so it is analogous to the weighting factors in Eq.~\ref{eq:GammaY}.
The difference is that $\hat Y^\text{rad}$ builds this weighting into the filter function whereas we opt to carry out the weighting when combining pixels.
The factor of $\bar H(f)$ in $\tilde Q^\text{rad}_{IJ}$ is the expected source power spectrum.
When we choose a set of pixels $\Gamma$, we effectively define $H(f)$ such that $H(f)=\mathrm{const}$ inside $\Gamma$ and $H(f)=0$ outside $\Gamma$.

Second, we note that apparently $\tilde Q^\text{rad}_{IJ}\propto\epsilon$ whereas our filter scales like $\tilde Q_{IJ}\propto 1/\epsilon$.
It turns out that both filters scale like $1/\epsilon$ because the radiometer normalization factor $\lambda\propto\epsilon^{-2}$.
The historical reason for this is that the radiometer analysis was developed by analogy with isotropic analyses~\cite{allen-romano}, which includes 
an integral over all sky directions.
The inclusion of $\gamma(t;f,\hat\Omega)$ in the expression for $\tilde Q_{IJ}^\text{rad}$ serves to weight different directions as more or less important just like the factor of $1/P_I(f)P_J(f)$ weights different frequencies.
We opt to remove $\gamma$ in favor of $\epsilon$, which deemphasizes the analogy with the isotropic analysis in order to highlight the network sensitivity, which is characterized by $\epsilon$.

\subsection{Relation to other search frameworks}
This is not the first time that $ft$-maps of data have been proposed to search for GWs.
The literature on this subject is extensive and diverse.
We concentrate on comparison with ``excess power'' methods, (see e.g.,~\cite{box,sylvestre,kalmus}).
The key difference between our framework and others
is that we cross-correlate data from two interferometers 
{\em before} they are rendered as $ft$-maps. 
Previous implementations such
as~\cite{sylvestre} and~\cite{kalmus} instead form $ft$-maps
 by auto-correlating data from each interferometer individually
and then correlating regions of significance in these maps.
For Gaussian noise, neither of these ways of
combining data from different detectors is optimal.
Instead, the optimal
multi-detector method incorporates both autocorrelated and cross-correlated
components~\cite{box}.
Real interferometric GW data, however, is not
Gaussian.
Rather, there is an underlying Gaussian component with frequent non-Gaussian bursts called ``glitches.''
For situations of this type, our approach has two advantages.

First, noise bursts in both detectors that coincide in time and frequency
increase the false-alarm rate for statistics with auto-correlated components,
but are suppressed in our cross-correlation analysis unless the waveforms of
the burst themselves are correlated in phase like a true GW.
Second, even when noise bursts are present, the pixel
values in an $ft$-map of cross-correlated data are well
approximated by a simple model. This is unlike $ft$-maps
with auto-correlated components, for which there is a no simple
description.
Thus, while our statistic is suboptimal for Gaussian data, we expect it to
perform well for real interferometer data. 
Moreover, even in the case of
Gaussian noise, we do not sacrifice much sensitivity compared to the optimal excess-power statistic, or
even to matched filtering, as demonstrated in Sec.~\ref{comparison}.
We compare the sensitivity of the cross-correlation statistic to other methods in Sec.~\ref{comparison}.

\section{Distribution of signal and background}\label{gaussianity}
In order to determine if a candidate event warrants further examination, it is necessary to determine the threshold above which an event is elevated to a GW candidate.
This threshold is usually phrased in terms of a false-alarm rate (FAR).
In Sec.~\ref{stats}, we argued that $ft$-maps of cross-power provide a convenient starting point for searches for long transients because cross-correlation yields a reasonably well-behaved $\text{SNR}(t;f,\hat\Omega)$ statistic whose probability density function (PDF) we can model numerically, thus allowing straightforward calculation of a nominal detection threshold in the presence of Gaussian noise.
We now assess this claim quantitatively.

We consider $\unit[52]{s}\times\unit[0.25]{Hz}$ pixels (created through a coarse-graining procedure described below), which are the intermediate data product in stochastic analyses such as LIGO's recent isotropic result~\cite{stoch-S5}.
There are, of course, other choices of pixel resolution, and different sources call for different resolutions.
Typically one must balance concerns about the signal duration, the signal bandwidth and the stationarity of the detector noise.
The PDF of $\text{SNR}(t;f,\hat\Omega)$ for a single pixel crucially depends on details of the pixel size.
E.g., the PDF of $\text{SNR}(t;f,\hat\Omega)$ for coarse-grained pixels (described in App.~\ref{gauss_items}) is more nearly Gaussian-distributed since coarse-grained pixels are created by averaging over more than one frequency.
Our goal here, therefore, is not an exhaustive treatment.
Rather, we aim to assess the agreement of data with our model using one pixel size, and in doing so, demonstrate how this assessment can be carried out in general.

\begin{figure}[hbtp!]
  \includegraphics[width=3.25in]{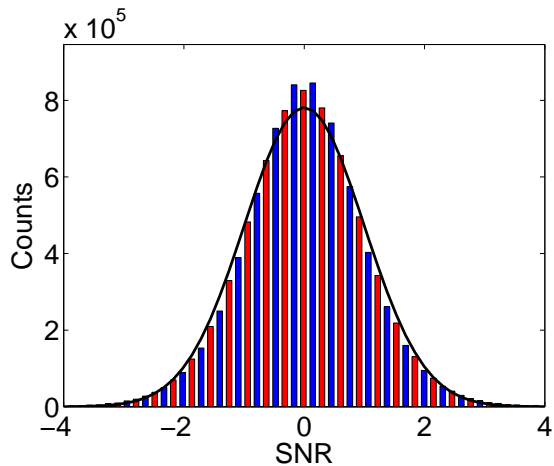} 
  \caption{Histogram of $\text{SNR}(t;f,\hat\Omega)$ using $\unit[52]{s}\times\unit[0.25]{Hz}$ pixels comparing S5 data with an unphysical time-shift (blue) to Monte Carlo data (red) and a normal distribution with $\text{mean}=0$ and $\sigma=1$ (black).  Error bars are too small to see.
    \label{fig:snr_pdf}}
\end{figure}

Our noise model assumes Gaussian strain noise, uncorrelated between detectors $I$ and $J$,
\begin{equation}
  2 \, \langle \tilde n_I(t;f) \tilde n_J(t;f) \rangle =
  \delta_{IJ} N_I(t;f) ,
\end{equation}
where $N_I(t;f)$ is the one-sided noise power spectrum and $\tilde n_I(t;f)$ is the discrete Fourier transform of the noise strain time series in detector $I$.
Although we are dealing with Gaussian noise, i.e., $\tilde n(t;f)$ is normally distributed, the associated PDF for $\text{SNR}(t;f,\hat\Omega)$ is {\em not} expected to be normally distributed.
It is more peaked than a normal distribution and it has broader tails (see Fig.~\ref{fig:snr_pdf}).

Fig.~\ref{fig:snr_pdf} shows a comparison of $\text{SNR}(t;f,\hat\Omega)$ for data and Monte Carlo.
The data corresponds to approximately a third of a day of at the beginning of LIGO's S5 science run using the Hanford H1 and Livingston L1 detectors.
We introduce an unphysical time-shift between the two data streams to remove all astrophysical content.
Additional data processing details are described in Apps.~\ref{gauss_items} and~\ref{gauss_app}, as the precise shape of the PDF for $\text{SNR}(t;f,\hat\Omega)$ depends crucially on details of how time series data is processed.

The Monte Carlo histogram is scaled by a normalization factor (derived analytically in App.~\ref{gauss_app}), which takes into account data processing not included in our Monte Carlo simulation, e.g., coarse-graining.
After applying this normalization factor, we find that the standard deviation of the data and Monte Carlo distributions agree to better than four significant digits.
We conclude that data and Monte Carlo are in qualitative agreement.
Thus we expect that the data are well-behaved enough that we can use a Gaussian noise model to assign a detection candidate threshold for $\text{SNR}_\Gamma$, at least for this choice of pixel size.

\section{Pattern recognition}\label{patterns}
In this section we showcase the cross-power statistic developed in Sec.~\ref{stats} using two different implementations (the box search and the Radon search) designed to address two different type of astrophysical scenarios (broadband signals and narrowband signals).

\subsection{Broadband box search}
\begin{figure}[hbtp!]
  \includegraphics[width=3.25in]{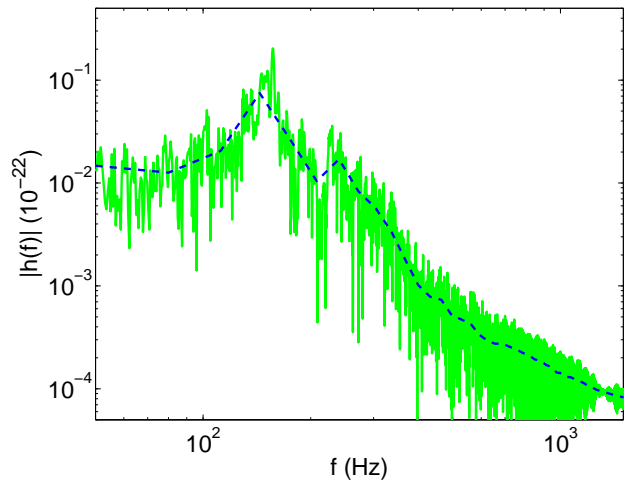} 
  \caption{GW strain amplitude spectrum 
    due to PNS convection in an axisymmetric PNS
    model at a typical galactic distance of $\unit[10]{kpc}$~\cite{ott:09}. 
    This plot was generated using the data simulated in~\cite{ott:09} available at~\cite{PNS_time_series}.
    \label{fig:PNS_spect}}
\end{figure}

\begin{figure*}[hbtp!]
  \begin{tabular}{cc}
    \includegraphics[width=2.85in,height=2.25in]{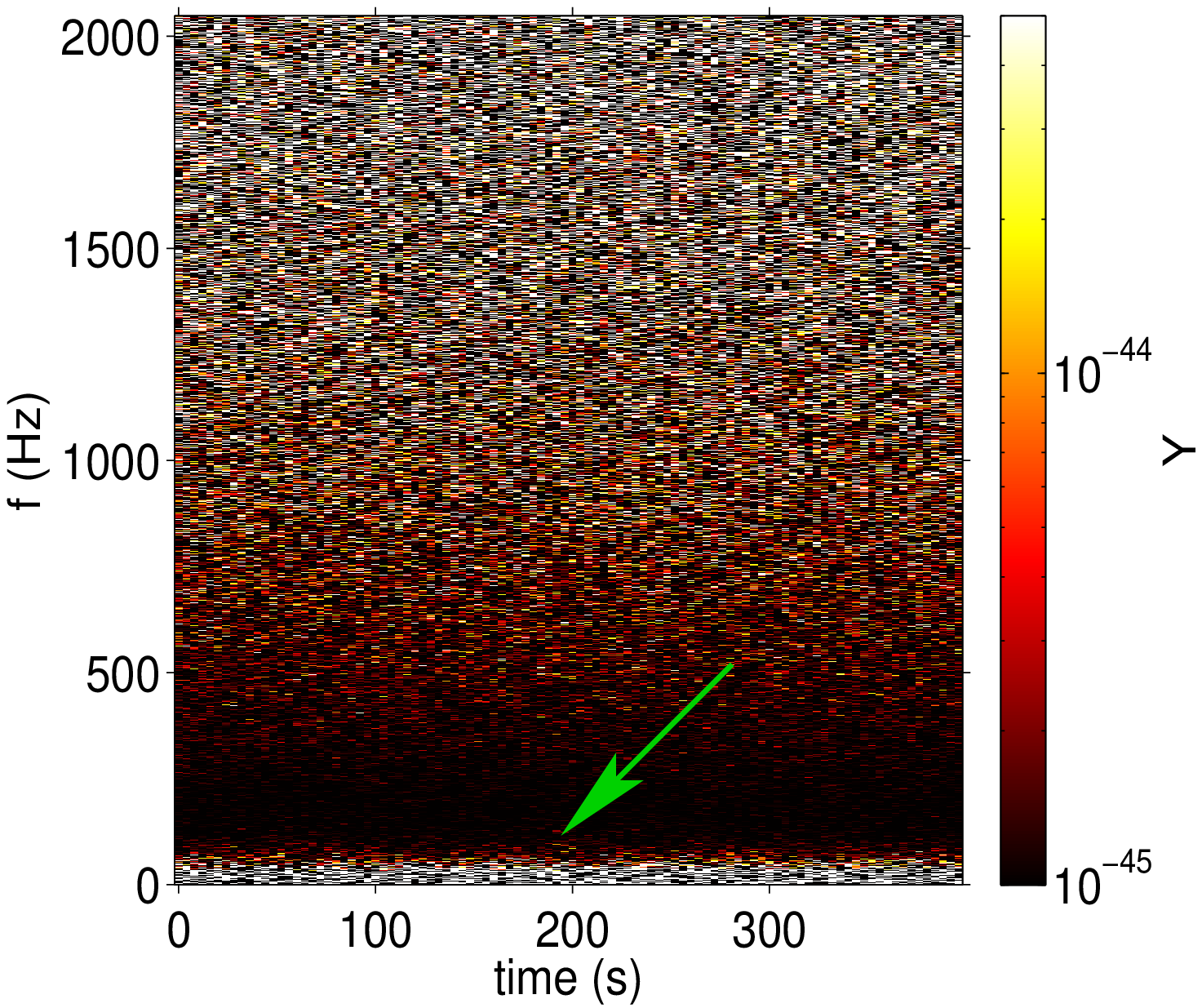} & 
    \includegraphics[width=2.85in,height=2.25in]{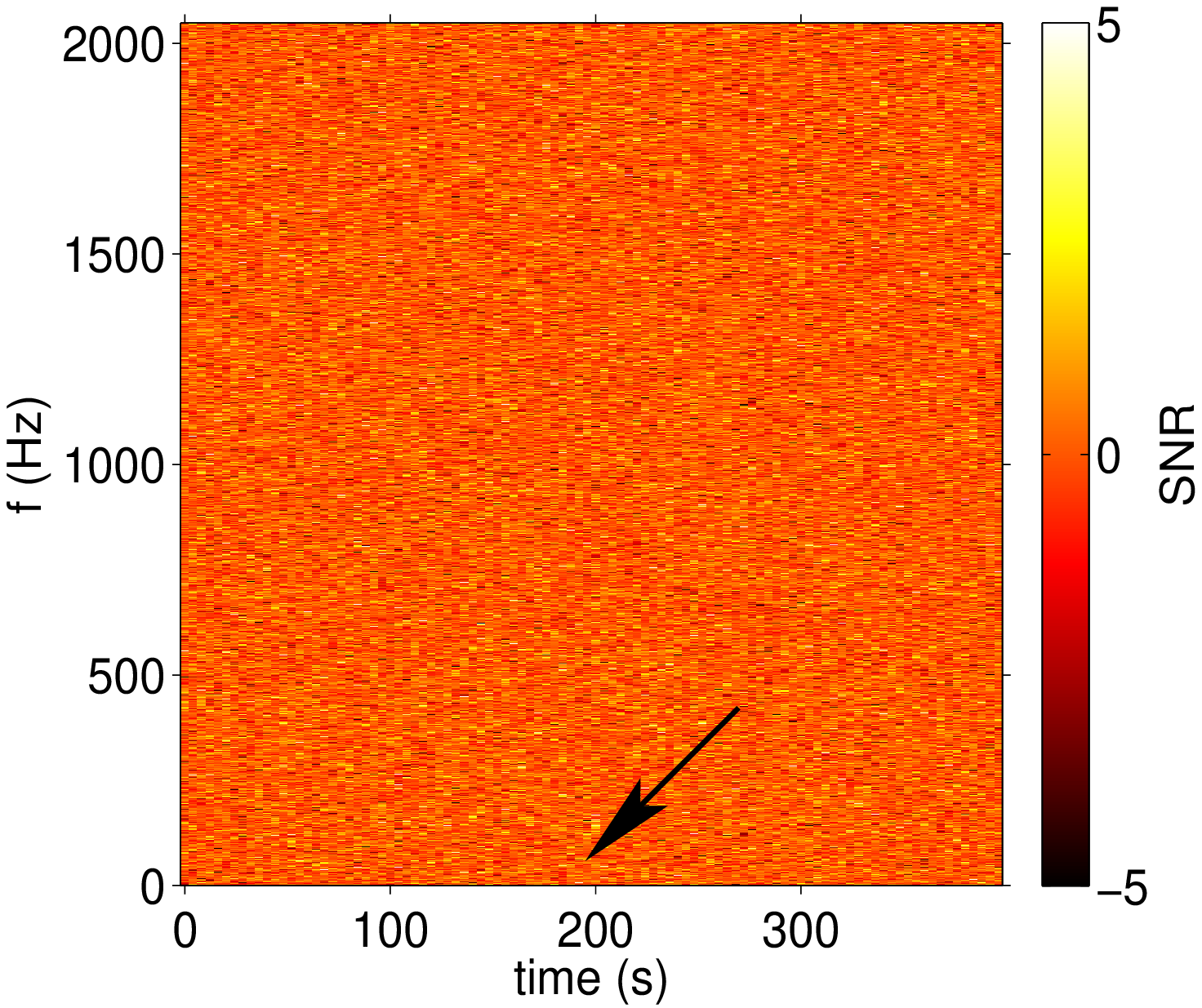} \\ 
    \includegraphics[height=2in]{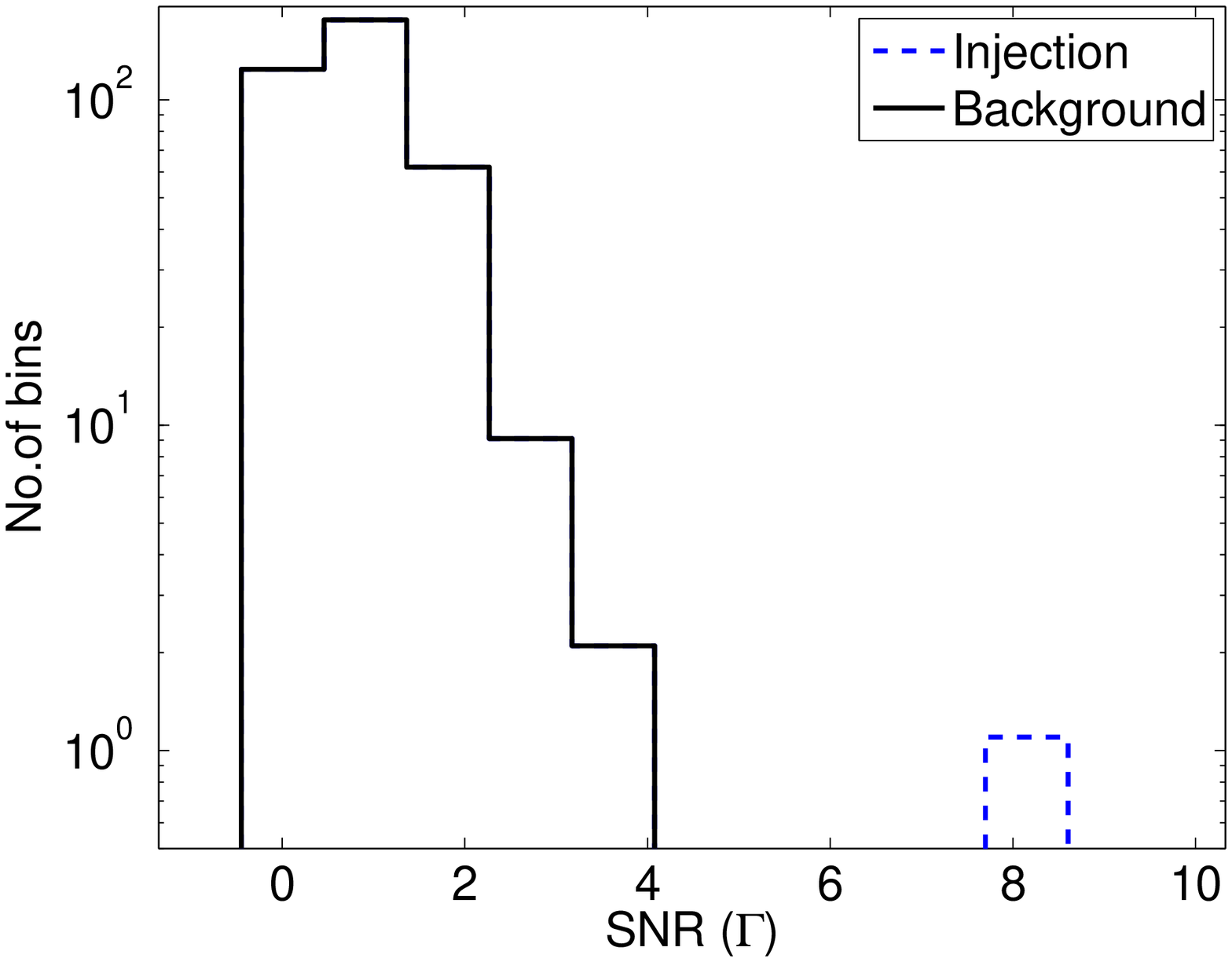} & 
    \includegraphics[height=2.1in]{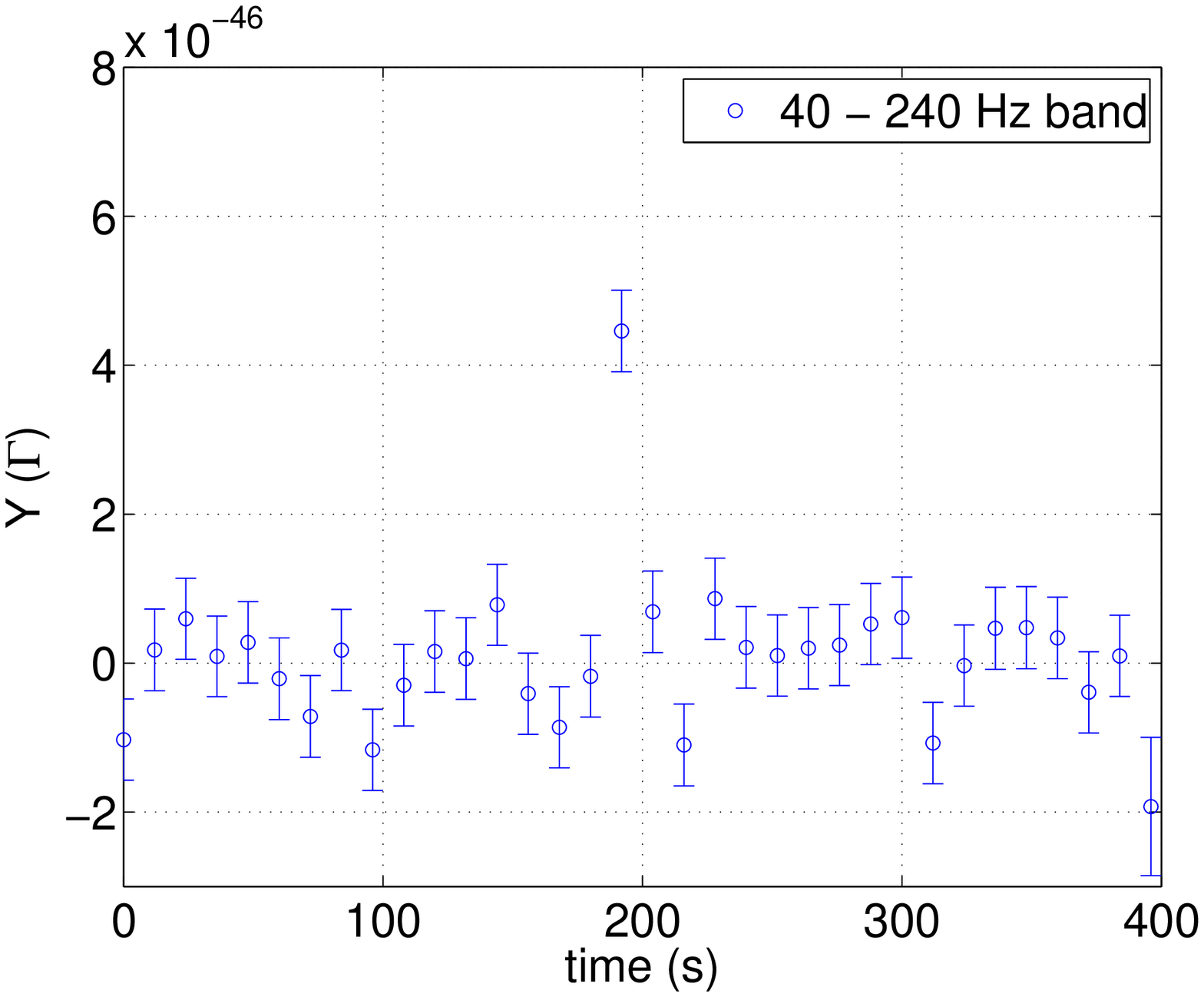} 
  \end{tabular}
  \caption{Injection recovery with the box-search algorithm. Top-left: an
    $ft$-map of $\hat Y(t;f,\hat\Omega)$.  
    The injected signal (not visible by eye) is indicated with a green arrow.
    Top-right: an $ft$-map of $\text{SNR}(t;f,\hat\Omega)$.
    The injected signal (not visible by eye) is indicated
    with a black arrow.  Bottom-left: a histogram of $\text{SNR}_\Gamma$ for
    a $\unit[200]{Hz}\times\unit[12]{s}$ box.
    The blue dashed corresponds to the injection.
    Though the signal is weak in each pixel, the signal obtained by 
    combining every pixel in $\Gamma$ is large.
    Bottom-right: $\hat{\cal Y}_\Gamma$ as a function of time.
     \label{fig:pns_inj}}
\end{figure*}

We demonstrate how a box-shaped set of pixels can be
used to search for a broadband GW transient source.
For illustrative purposes,
we consider a toy model based on protoneutron star (PNS) convection
with a spectrum produced in an axisymmetric PNS model assuming a non-rotating, 
$15$ $M_{\astrosun}$ progenitor~\cite{ott:09} (see Fig.~\ref{fig:PNS_spect}).
We simulate a $d=\unit[4.5]{kpc}$
source in the direction of $\text{ra}=\unit[17]{hrs}$, $\text{decl}=30^{\circ}$ at 00:00~GMST on top of simulated detector noise comparable to the design sensitivity for initial LIGO.
We calculate the cross-power statistic $\hat{\cal Y}_\Gamma$ utilizing a $\unit[200]{Hz}\times\unit[16]{s}$ box constructed with the H1L1 detector network.
We use $\unit[4]{s}\times\unit[0.25]{Hz}$ pixels, and for each pixel we use $20$ adjacent segments to calculate $\hat\sigma(t;f,\hat\Omega)$, ($10$ on each side).
We tile the $ft$-map and record the $\hat{\cal Y}_\Gamma$ within each box.
We find that the signal can be recovered with $\text{SNR}(t;f,\hat\Omega)=8$.
The results are summarized in Fig.~\ref{fig:pns_inj}.

\subsection{Radon algorithm}\label{radon}
Radon transforms are regularly used in imaging problems in order to
identify line-like features in 2D maps~\cite{radon_transform}.
This makes the Radon algorithm useful for looking for narrowband GW tracks in $ft$ maps, (see, e.g.,~\cite{LISA_Radon}).
By converting from the
coordinates $(t;f)$ to impact parameter $b$ and angle $\theta$, a line-like cluster in $ft$-space is converted to a peak in Radon space.
Thus, this algorithm provides a convenient way to search for GW signals that manifest themselves as line-like tracks in $ft$-space.

For continuous variables, the Radon transform of some function $g(t;f)$ is defined as~\cite{radon_transform}:
\begin{equation}
  \mathcal{R}(b,\theta)[g(t;f)] \equiv \int df \int dt \, 
  g(t;f) \delta(b -t\cos\theta - f\sin\theta) .
\end{equation}
For discrete variables, the Radon transform becomes
\begin{equation}\label{eq:RadonDefn}
  \mathcal{R}(b,\theta)[g(t;f)] = \sum_{t;f} w_{t;f}^{b\theta} \, g(t;f) .
\end{equation}
The weight factors $w_{t;f}^{b\theta}$ describe how close a line, parameterized by $(b,\theta)$, passes to the center of each $ft$-pixel.
We use a (modified) Radon transform algorithm from~\cite{myradon}, which is one of many possible implementations of the discrete Radon transform.

The estimator for the cross power in a pixel set described by $(b,\theta)$ can be written entirely in terms of Radon transforms:
\begin{equation}\label{eq:YRadon}
  \hat{\cal Y}(\hat\Omega,b,\theta)
  = \frac{ 
    \mathcal{R}[\hat Y(t;f,\hat\Omega) \hat\sigma(t;f,\hat\Omega)^{-2}]
  }{
    \mathcal{R}\left[\hat\sigma(t;f,\hat\Omega)^{-2}\right]
  }
\end{equation}
The associated variance is
\begin{equation}\label{eq:SigmaGammaYRadon}
  \begin{split}
    \hat\sigma_{\cal Y}(\hat\Omega,b,\theta)^2 & =
    \frac{
      \sum_{t;f} (w_{t;f}^{\theta b})^2 \, \hat\sigma(t;f,\hat\Omega)^{-2} 
    }{
      \left( \mathcal{R}[\hat\sigma(t;f,\hat\Omega)^{-2}] \right)^2 
    }
    \end{split}
\end{equation}

\begin{figure}[hbtp!]
  \begin{tabular}{c}
    \includegraphics[width=3.25in]{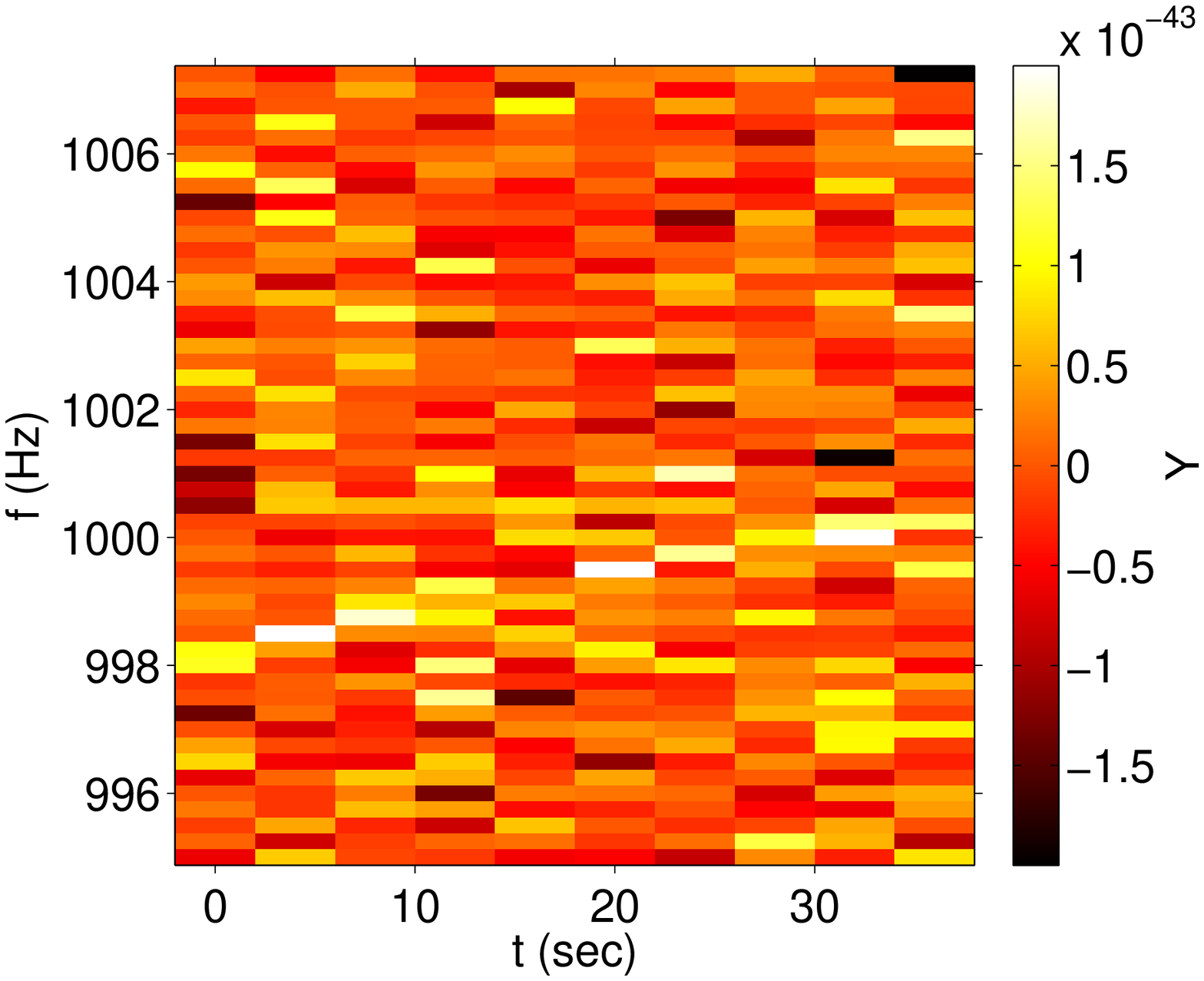} \\ 
    \includegraphics[width=3.25in]{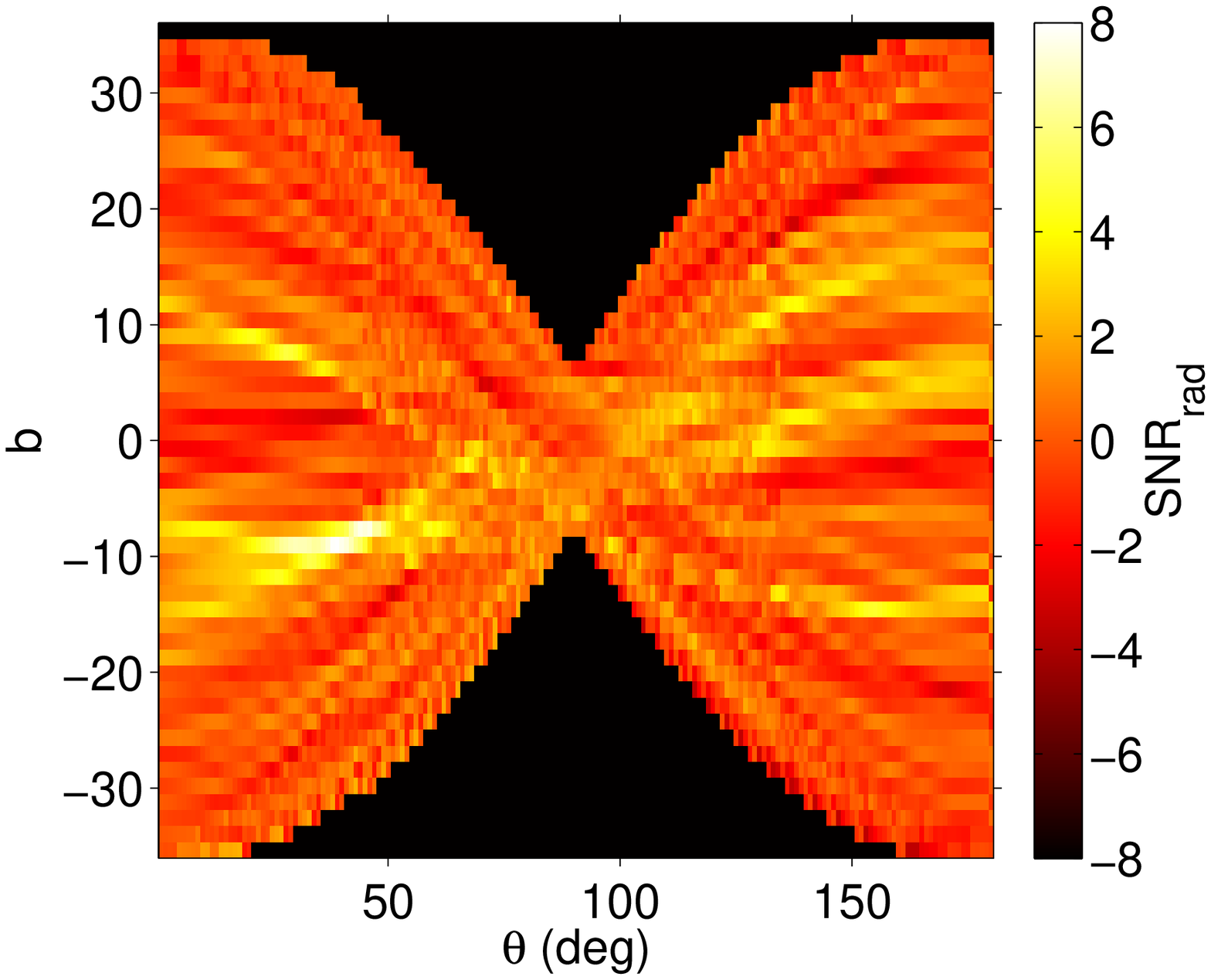} 
  \end{tabular}
  \caption{
    Top: an $ft$-map of $\text{SNR}(t;f,\hat\Omega)$.  
    The long-GRB track is a barely discernible diagonal line.
    Bottom: a map
    of $\text{SNR}_\Gamma(\hat\Omega,b,\theta)$.  The signal
    shows up as a hot spot at
    $(\theta,b)=(39^\circ,-8)$ with
    $\text{SNR}_\Gamma=8.1$. The sinusoidal patterns are (expected)
    covariances between Radon-map pixels. \label{fig:vPYmap}}
\end{figure}

We now consider a toy model of torus excitations from long GRBs~\cite{vanPutten}, which are expected to produce line-like clusters in $ft$-space with durations of $\unit[2\sim200]{s}$.
Since we are dealing with an elliptically polarized source, $\hat{\cal Y}(t;f,\hat\Omega)$ also depends on inclination angle $\iota$ and polarization angle $\psi$, (see App.~\ref{polarized}).
For the sake of simplicity, however, we use an unpolarized filter, which has been shown to do a reasonably good job recovering elliptically polarized sources~\cite{stefan}.
We simulate an elliptically polarized waveform (see Tab.~\ref{tab:vP_params}) on top of simulated detector noise comparable to design sensitivity for initial LIGO.
Once again, we use $\unit[4]{s}\times\unit[0.25]{Hz}$ pixels, and for each pixel we use $18$ adjacent segments to calculate $\sigma(t;f,\hat\Omega)$ ($9$ from each side).
The resulting maps of $\hat Y(t;f,\hat\Omega)$ and $\text{SNR}_\Gamma$ are shown in Fig.~\ref{fig:vPYmap}. For this toy-model, the signal was recovered with $\text{SNR}_\Gamma=8.1$ at $d=\unit[1.7]{Mpc}$.

\begin{table}[tp]
  \begin{tabular}{ll}
    \toprule
    parameter & value \\
    \toprule
    strain amplitude & $h_\text{rms}=1.2\times10^{-21}$ \\
    search duration & $\unit[100]{s}$ \\
    (ra, dec) & $(\unit[17]{hr},30^\circ)$ \\
    $(\iota, \psi)$ & $(0^\circ, 0^\circ)$ \\
    \toprule
    injection start time & 00:00 GMST \\
    injection duration & $\unit[40]{s}$ \\ 
    distance to GRB & $\unit[1.7]{Mpc}$ \\
    $(f_0, df/dt)$ & $(\unit[998]{Hz}, \unit[0.03]{Hz/s}$) \\
    \toprule
  \end{tabular}
  \caption{Search and injection parameters used for Fig.~\ref{fig:vPYmap}.
    \label{tab:vP_params}}
\end{table}

\subsection{Other algorithms}
In this section we have, for illustrative purposes, presented two of the many pattern recognition algorithms that may be applied to the problem of looking for features in $ft$-maps of cross-power.
There is a diverse and extensive literature devoted to the study of cluster identification, (see, e.g.,~\cite{honda,kahn}).
In the next section we apply the Radon algorithm to non-GW channels in order to look for environmental noise artifacts that are qualitatively similar to our long-GRB toy model.
For comparison, we also make use of locust and Hough algorithms~\cite{raffai}, which have been proposed as a method of identifying long-GRB events in GW data.

\section{Application to environmental noise identification}\label{pem}
\subsection{Environmental noise in GW interferometers}
While our discussion until now has been focused on the detection of GW transients, the same formalism can be applied to look for structure in $ft$-maps of cross-power between any two data channels.
In particular, it is illuminating to study the cross-power between an interferometer's GW-strain channel, (which we denote $s_\text{GW}$) and a physical environmental monitoring (PEM) channel such as a seismometer or a magnetometer channel located near the interferometer.
Since PEM channels are not sensitive to GWs, statistically significant features in an $ft$-map of PEM-$s_\text{GW}$ cross-power are likely due to environmentally-induced noise artifacts.

Transient artifacts are called ``glitches'' whereas persistent narrowband features are often called ``lines'' or ``wandering lines'' when the frequency slowly changes over time.
Glitches and wandering lines can be problematic for searches for bursts / compact binary coalescences and for pulsars respectively, see, e.g.,~\cite{cbc_glitches,S6DetectorChar,Einstein,PowerFlux}.
(They also produce non-Gaussian noise for our cross-power statistic.)
It is thus desirable to identify and when possible mitigate these noise features.

In this section we show how the formalism we have developed to search for long GW transients can also be used to identify glitches and wandering lines in PEM-$s_\text{GW}$ cross-power maps.
There are two points we hope to make with this digression.
First, we shall see that PEM-$s_\text{GW}$ $ft$-maps are useful for identifying, characterizing (and in some cases eliminating) environmentally-induced noise.
Second, we show that some environmentally-induced noise in PEM-$s_\text{GW}$ $ft$-maps is qualitatively similar to the GW-transient signature in cross-power maps between two $s_\text{GW}$ channels.
Thus, these noise artifacts provide a convenient dataset to demonstrate our search algorithms with environmental noise events.

\subsection{Environmental channels at LIGO and Virgo}
In order to facilitate the detection of transient GWs, it is necessary to monitor and characterize glitches and lines.
Efforts to identify and document noise artifacts are a major task of the Detector Characterization and Glitch groups within the LIGO Scientific Collaboration~\cite{Blackburn, S6DetectorChar,Fscan} and the Virgo Collaboration.
To assist in this effort, each LIGO/Virgo observatory is supplemented with hundreds of sensors that monitor the local environment.
(For an overview of the LIGO and Virgo interferometers, see~\cite{ligo_ropip} and~\cite{VIRGO} respectively.)

Accelerometers measure vibrations such as the motion of the beam tubes and of the optical tables that house photodiodes; microphones monitor acoustic noise at critical locations; magnetometers monitor magnetic fields that could couple to the detector; radio receivers monitor radio frequency power around the laser modulation frequencies; and voltage line monitors record fluctuations in the AC power.
The PEMs are placed at strategic locations around the observatory, especially near the corner and ends of the interferometer where important laser, optical and suspension systems reside in addition to the test masses themselves.

For illustrative purposes, we consider a special class of noise artifacts induced by passing airplanes.
These ``airplane events,'' have attractive properties for our purposes.
First, airplane events are relatively well understood.
The existing LIGO airplane veto system (called {\tt planemon}) has been shown to flag airplanes observed in microphone channels, and these flags have been shown to agree with airplane flight data~\cite{planemon,riles}, though the existing {\tt planemon} algorithm does not determine if the passing airplane affects $s_\text{GW}(t)$.
Since we already understand a lot about airplane events, it is straightforward to assess if our algorithms are consistent with what we already know.

Second, we shall see that airplanes produce a slightly curved narrowband $ft$-map track lasting tens of seconds.
These tracks are qualitatively similar to GW transient scenarios such as van Putten's long-GRBs model, and thus they provide an opportunity for us to demonstrate our search algorithms on a distribution of signals that resemble unmodeled GW transients; we have a qualitative picture of the signal, but it is impractical to model the space of all possible signals with a matched filter template bank.

\subsection{Airplane noise identification}
In order to identify airplane events, $ft$-maps correlating the GW channel with acoustic channels are computed in $\unit[400]{s}$ blocks.
Then we take the absolute value of the $\text{SNR}(t;f)$ map.
This step is not necessary for GW studies because GW transients can only produce positive-definite $\text{SNR}(t;f)$ once the phase delay between two sites has been taken into account.
Transient noise artifacts, on the other hand, can produce complex (not positive-definite) $\text{SNR}(t;f)$ since the cross-power phase depends on the coupling of the environmental noise into $s_\text{GW}$.
By taking the absolute value of $\text{SNR}(t;f)$, the PDF of $\text{SNR}(t;f)$ changes from the description in Subsec~\ref{gaussianity}, so we estimate it semi-empirically with the assistance of simulation (see below).

\begin{figure*}[hbtp!]
  \begin{tabular}{cc}
    \includegraphics[width=3.25in]{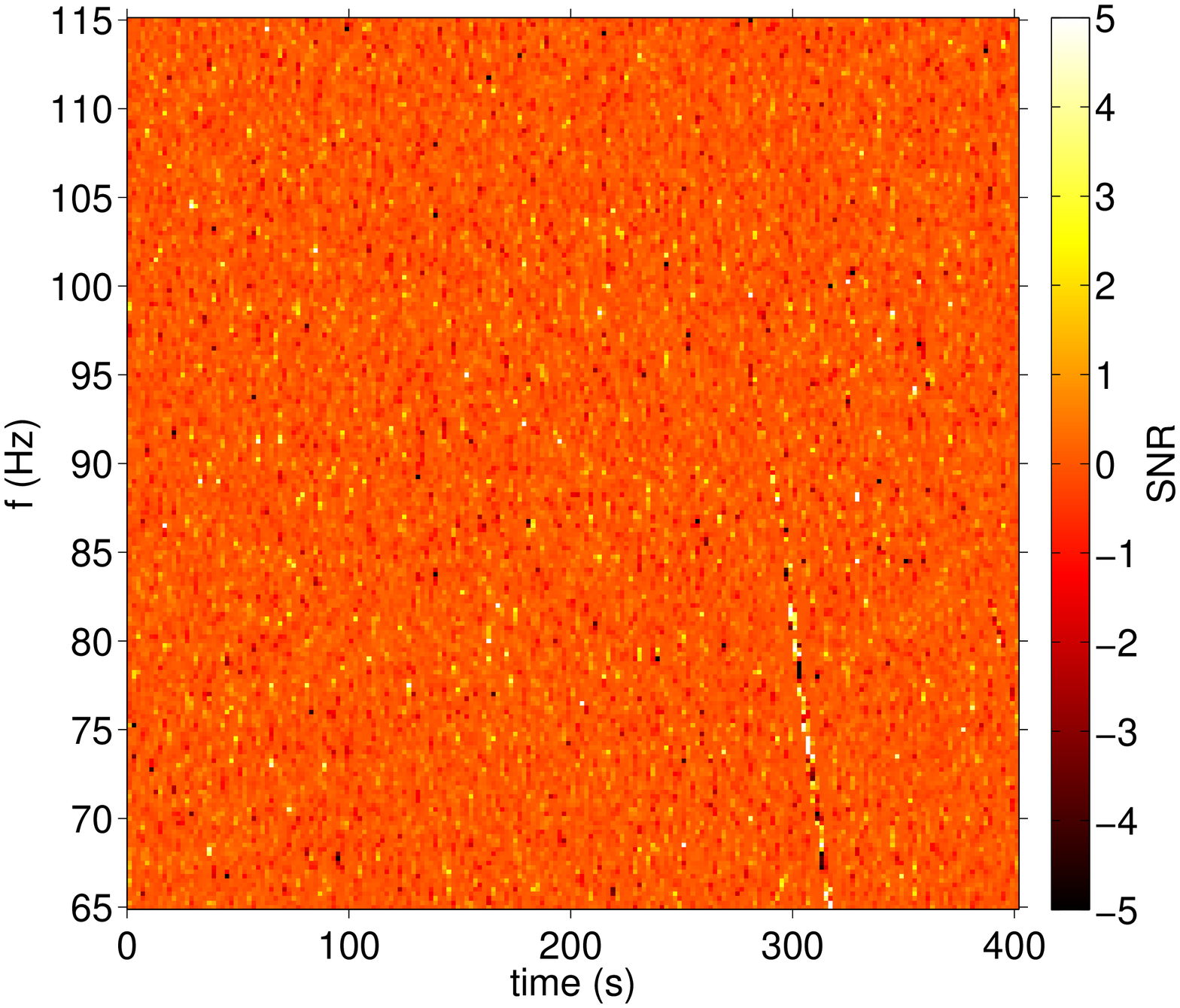} & 
    \includegraphics[width=3.25in]{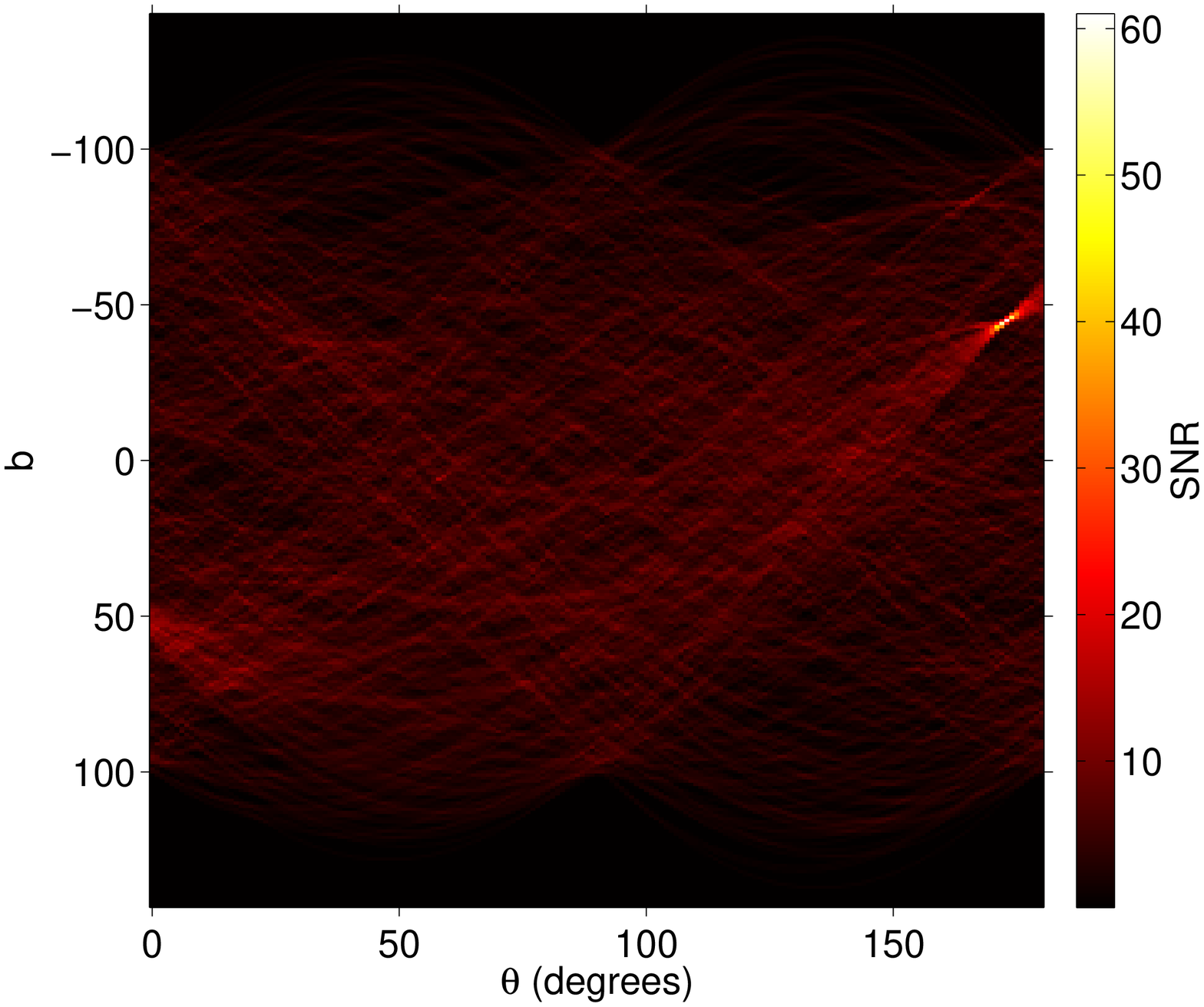} \\ 
    \includegraphics[width=3.25in]{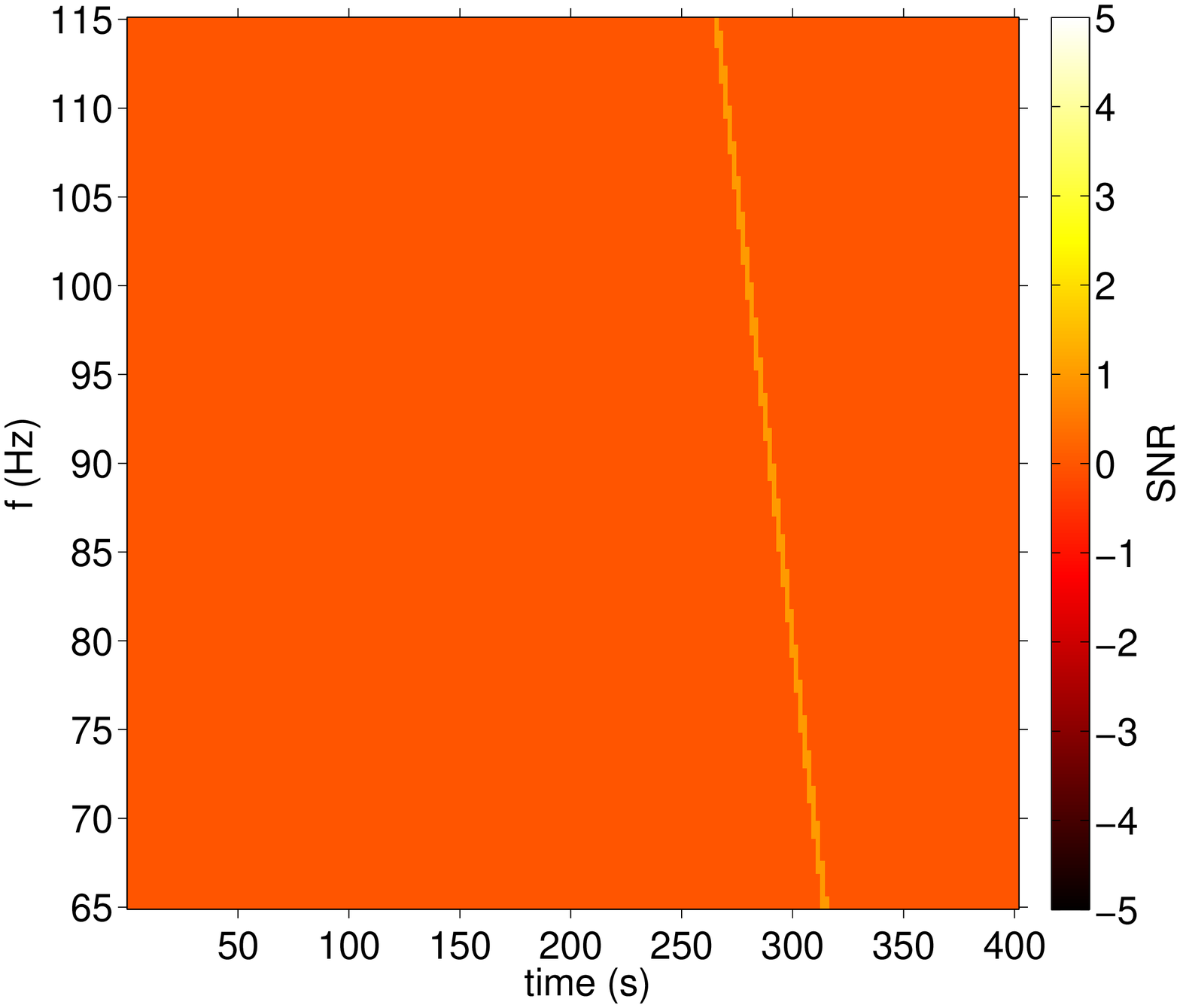} & 
    \includegraphics[width=3.25in]{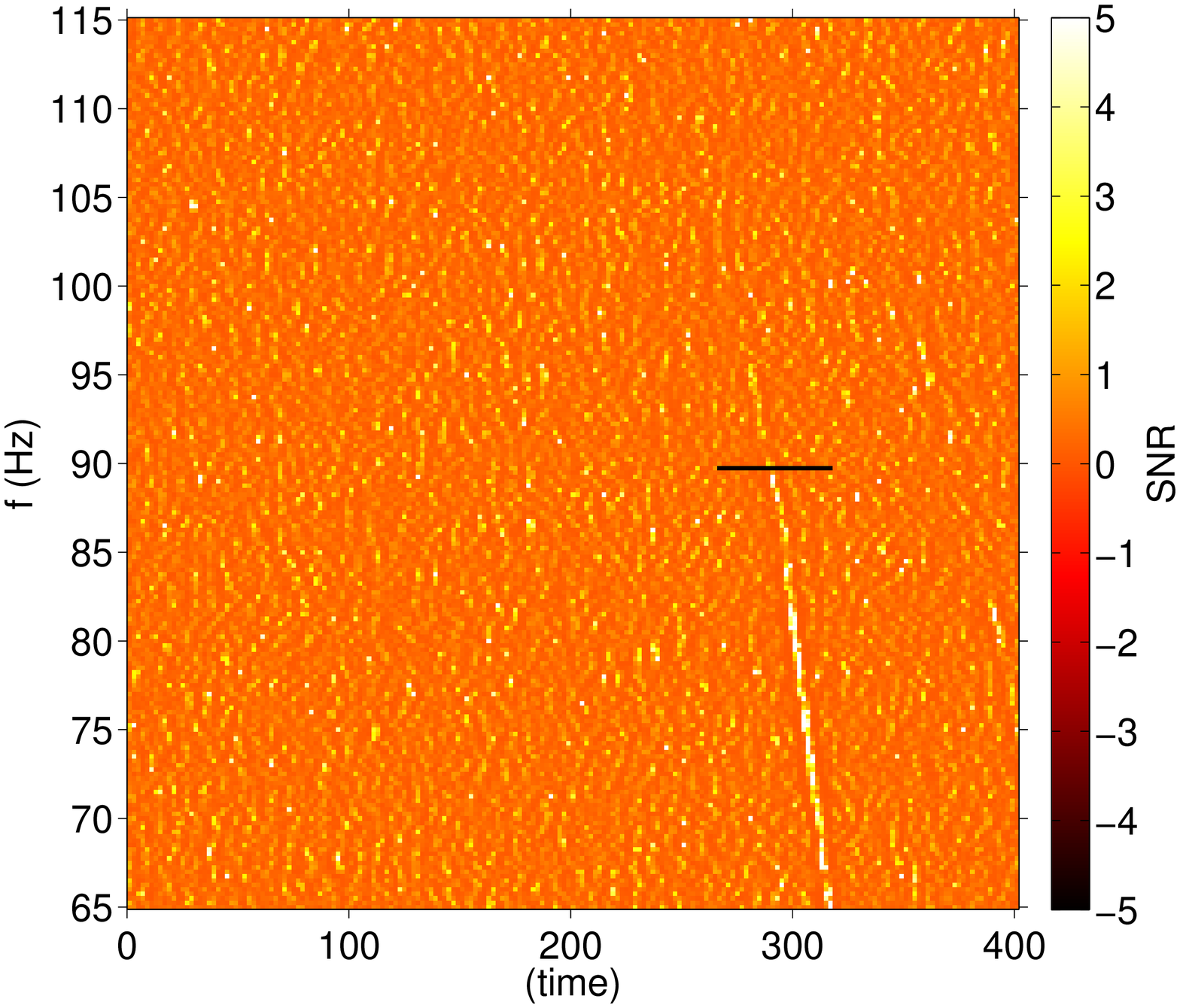} 
  \end{tabular}
  \caption{Top-left: a $\unit[400]{s}$-duration map of $\text{SNR}(t;f)$ 
    (see Eq.~\ref{eq:pixelSNR})
    created with $\unit[4]{s}\times\unit[0.25]{Hz}$ pixels and 
    using $s_\text{GW}$ cross-correlated with a microphone.  The
    slightly curved track on the right side of the plot is caused by
    the Doppler-shifted acoustic signal from a passing airplane.
    Top-right: the associated Radon map. Note the bright spot on the
    mid-right corresponding to the airplane track.  Bottom-left:
    $ft$-map of the reconstructed track using the maximum $\text{SNR}(t;f)$
    pixel in
    Radon space.  Bottom-right: $ft$-map of the magnitude of $\text{SNR}(t;f)$
    including a black line corresponding to the veto
    window.  These data are from the beginning of LIGO's S5 science run. \label{fig:airplanes}}
\end{figure*}

\begin{figure}[hbtp!]
  \includegraphics[width=3.25in]{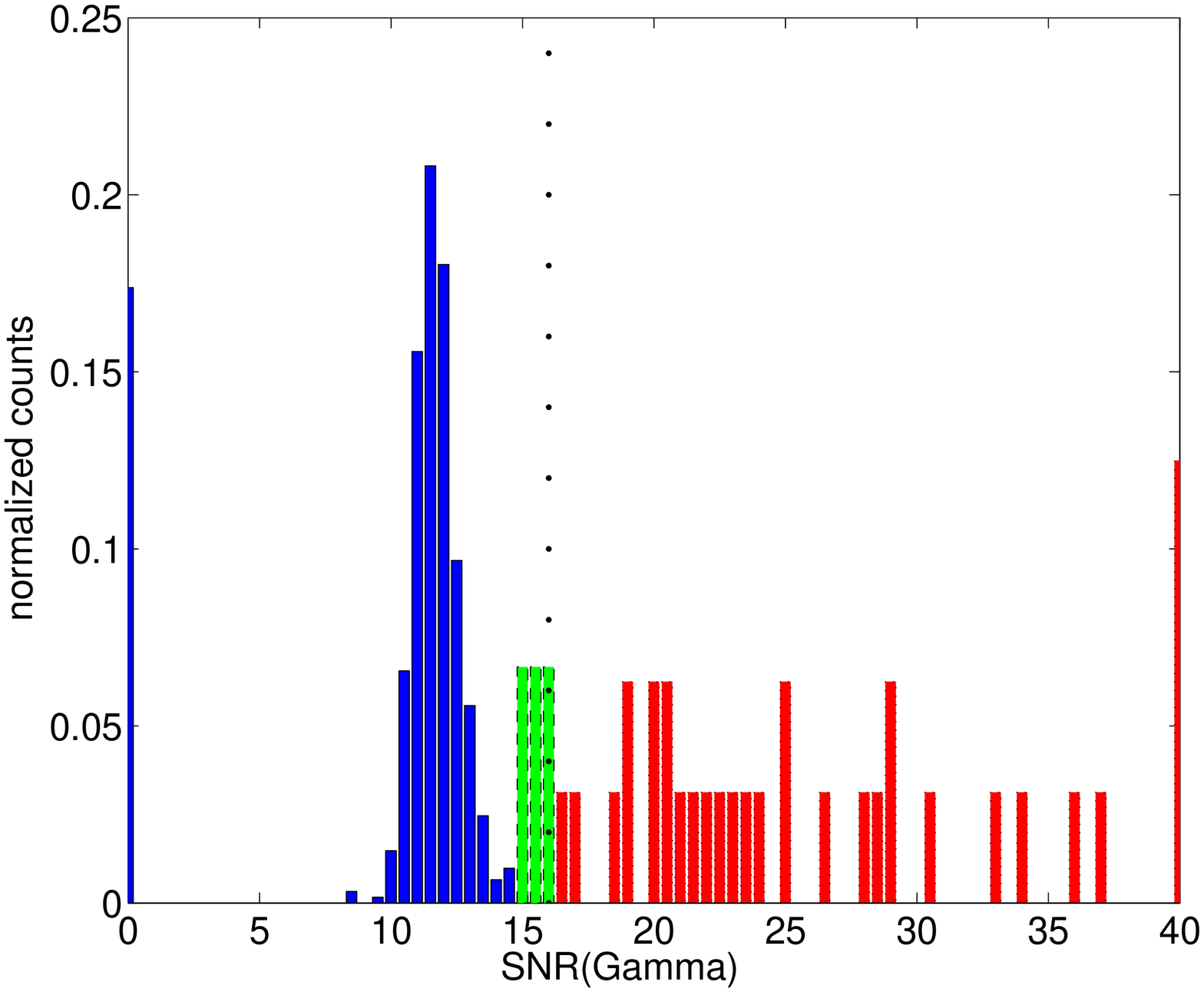} 
  \caption{A histogram corresponding to $\text{SNR}_\Gamma$ for
    above-threshold airplane events (in red, dotted outline),
    near-threshold events (green dashed outline) and below-threshold background
    events (in blue, solid outline) in three days of LIGO Hanford Observatory 
    data. 
    The dotted
    black line corresponds to the chosen threshold of 16. The red,
    green and blue distributions are separately normalized to unity
    for the purpose of plotting. Entries with
    $\text{SNR}_\Gamma=0$ record the maps which were excluded from this study
    for failing a ``glitch cut.''
    Entries with $\text{SNR}_\Gamma=40$ record the maps
    with $\text{SNR}_\Gamma\geq40$. \label{fig:Histogram}}
\end{figure}

We create a Radon transform of each $ft$-map of $|\text{SNR}(t;f)|$, an example of which can be seen in Fig.~\ref{fig:airplanes}.
We are presently interested in qualitative aspects of the airplane track, and so we make the simplifying assumption that $\sigma(t;f)=\mathrm{const}$.
Although airplane tracks are slightly curved, the approximation of the tracks as lines is suitable for a simple identification.
We record the brightest spot, i.e., the maximum $\text{SNR}_\Gamma$, on each Radon map.
The maps are then ordered according to their maximum $\text{SNR}_\Gamma$.
The $ft$-maps with the highest $\text{SNR}_\Gamma$ are checked by eye for airplane tracks.
In Fig.~\ref{fig:Histogram}, we show a histogram of $\text{SNR}_\Gamma$.
The red entries are for $ft$-maps with unambiguous airplane tracks.
The green entries are possible airplane tracks, yet a visual inspection of these events was inconclusive.
All other entries (due to non-airplane background) are blue.
Using Fig.~\ref{fig:Histogram}, we set a threshold of $\text{SNR}_\Gamma>16$ for an event to be tagged as an unambiguous airplane.

In Tab.~\ref{tab:airplane}, we compare the Radon triggers to the {\tt planemon} triggers.
During a span of $\unit[3]{days}$, the Radon algorithm found 37 unambiguous airplanes, 26 of which were also seen by {\tt planemon}.
The 11 events that were not confirmed by {\tt planemon} were confirmed by eye as airplane-like.
It is possible that they failed to create a {\tt planemon} trigger because the airplane flight path did not trigger enough microphone channels.
During this same period, {\tt planemon} found 54 unique events, 28 of which were not found by the Radon algorithm as the tracks were not coherent in $s_\text{GW}$.

\begin{table}[tp]
  \begin{tabular}{lcc}
  \toprule
     & {\tt planemon}-flagged & not {\tt planemon}-flagged \\
     Radon-flagged & 26 & 11 \\
     not Radon-flagged & 28 & 583 \\ 
    \toprule
  \end{tabular}
  \caption{Number of airplane event triggers identified by the Radon and {\tt planemon} algorithms out of 648 total $ft$-maps. \label{tab:airplane}}
\end{table}

We conclude that the Radon flags appear to be roughly consistent with the planemon flags.
However, the Radon flag differs in two useful ways.
First, Radon flags require coherence between microphone channels and $s_\text{GW}$, and so our algorithm only flags airplanes that contaminate $s_\text{GW}(t)$.
Second, the Radon flag does not need confirmation between more than one microphone channel so long as the signal from one is coherent with $s_\text{GW}$.

For each map deemed to contain an airplane, the inverse Radon transform is performed on the brightest spot, and the airplane track is reconstructed as an $ft$-map line as in Fig.~\ref{fig:airplanes}.
The start and stop times of the airplane noise in this frequency band are estimated to be the times at which the reconstructed track intersects the edge of the $ft$-map.
This routine was run on the LIGO and Virgo data during recent science runs, correlating the GW channel with about 10 microphones at each detector. The algorithm identified $\sim10-15$ airplane events at an observatory each day.

PEM-$s_\text{GW}$ $ft$-maps created with real data contain additional environmental noise artifacts besides airplanes.
In order to estimate the FAR for airplane tracks in ``idealized'' noise (where no other environmental artifacts are present), we perform Monte Carlo pseudo-experiments by scrambling $ft$-map pixels so as to wash out clustering while preserving the empirically observed distributions of $Y(t;f)$ and $\sigma(t;f)$.
Using the weighted Radon algorithm, (where $\sigma(t;f)\neq \mathrm{const}$), we estimate that, on average, $\unit[0.4]{events/month}$ are falsely identified as airplanes in idealized noise.

\subsection{Other algorithms}
Having demonstrated how the Radon algorithm can recover airplane events in PEM-$s_\text{GW}$ $ft$-maps, we now demonstrate two additional pattern recognition algorithms.
Our point is to convey the wide array of tools available to solve the problem of pattern recognition in $ft$-maps of cross-power.
There are typically both advantages and disadvantages associated with each algorithm, which means that each one lends itself to different applications, though extensive discussion of the merits of different techniques is beyond our current scope.

In particular, we consider the locust and Hough algorithms from~\cite{raffai}, both of which are well-suited for narrow-band sources.
The Hough algorithm is similar to the Radon algorithm, except that it can be extended to fit tracks described by arbitrarily high-order polynomials.
By introducing additional fit parameters, the tracks tend to be reconstructed more accurately.
However, by adding more parameters, the significance of a line-like event with little or no curvature can be less than the value obtained by the Radon algorithm.

The locust algorithm is a local wandering algorithm, which integrates the $ft$-map along a chain of local maxima.
This algorithm has the advantage that it can reconstruct arbitrary-shaped tracks without large numbers of free parameters.
Since it relies on local maxima, however, the Radon and Hough algorithms are more robust if the GW power is spread diffusely over many pixels.
Both the locust and Hough algorithms produce a statistic, which is the integral of cross-power along a track.
We estimate significance by performing Monte Carlo pseudo-experiments in which we randomly scramble the $ft$-map pixels.

\begin{figure}[hbtp!]
  \begin{tabular}{c}
  \includegraphics[width=3.25in]{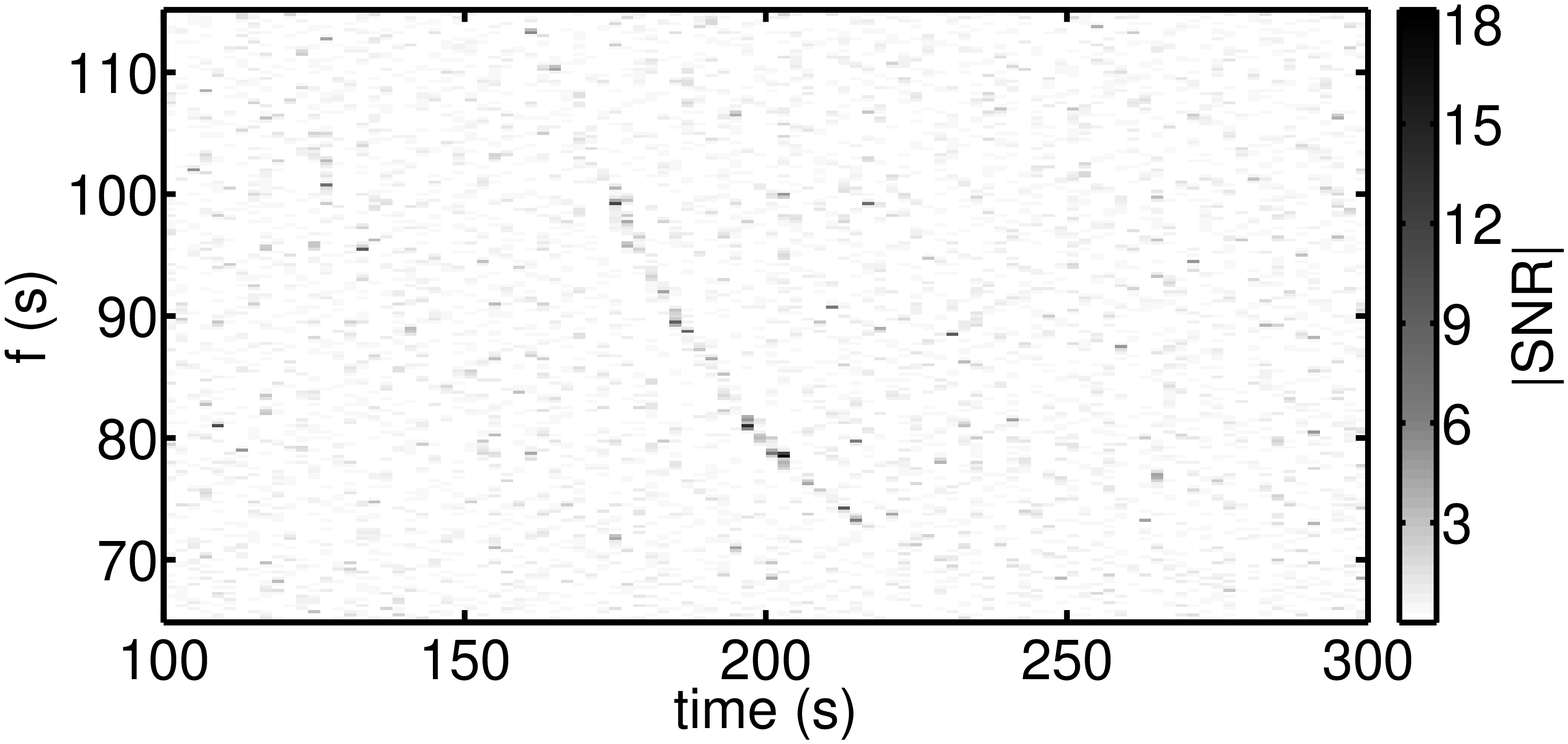} \\ 
  \includegraphics[width=3.25in]{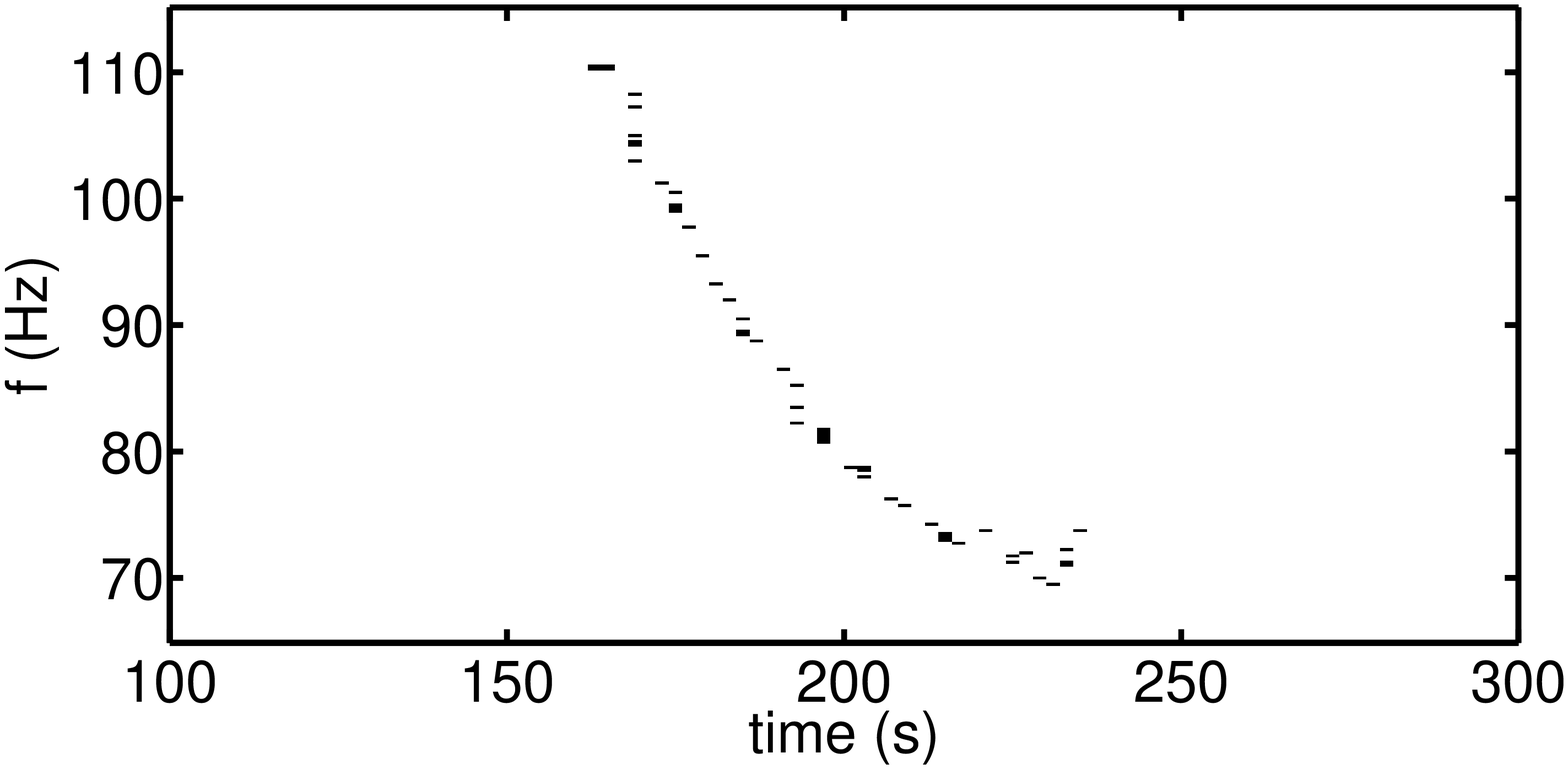} \\ 
  \includegraphics[width=3.25in]{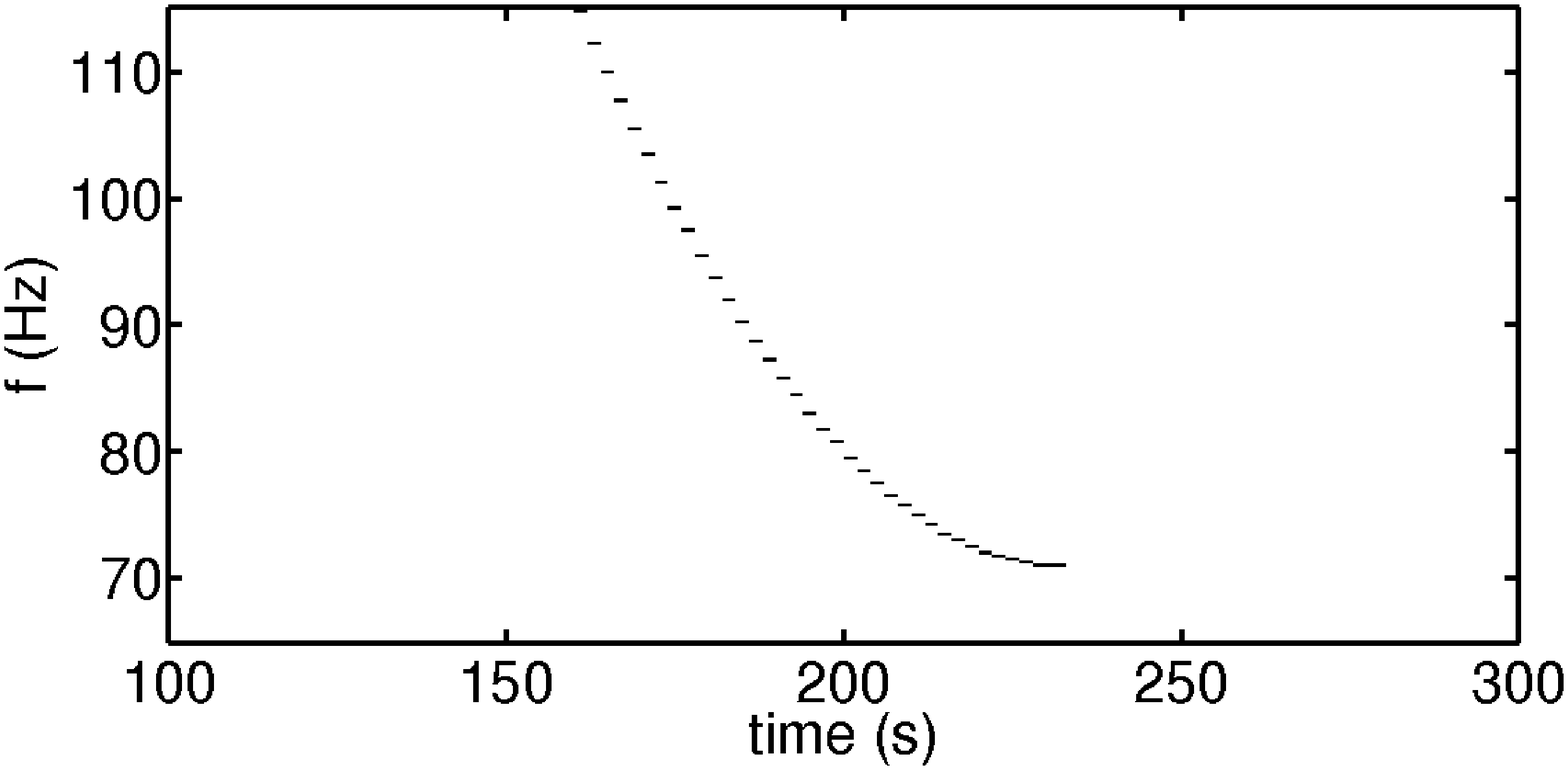} 
  \end{tabular}
  \caption{Upper panel: an airplane track in a map of
    $\text{SNR}(t;f,\hat\Omega)$.  Middle panel: the
    reconstructed track as determined by the locust algorithm.  Bottom
    panel: the reconstructed track as determined by the Hough
    algorithm using a second-order
    polynomial. \label{fig:locust_hough}}
\end{figure}

Applying the locust and Hough algorithms to an unambiguous airplane event, we obtain the reconstruction plots shown in Fig.~\ref{fig:locust_hough}.
We determine that both the locust and Hough algorithms detect the event with a FAR no more than $0.04\%$ per $\unit[400]{s}$ map in idealized noise.

\section{Comparison with other techniques}\label{comparison}
In this section, we compare the proposed excess cross-power statistic to matched filtering and to the general excess-power statistic from~\cite{box}.
It is impractical to actually carry out a matched filtering analysis since we do not have a template bank from which we can construct an arbitrary transient signal.
Nevertheless, it is possible to perform analytical estimates.

We consider a signal space characterized by an $ft$-volume $V$ spanned by $N_\text{eff}$ independent matched filtering templates.
The $ft$-volume is simply the number of pixels in our pixel set $\Gamma$.
We endeavor to address the following question: given a false-alarm probability (FAP) and a false-dismissal probability (FDP), what is the minimum signal amplitude detectable by either method?
Following~\cite{box}, we respectively define thresholds $A^\mathrm{CP}_\text{min}$, $A^\mathrm{EP}_\text{min}$ and $A^\mathrm{MF}_\text{min}$ as the minimum detectable amplitudes by our cross-power (CP) search, by the optimal total excess-power (EP) search~\cite{box}, (which includes both cross-power and auto-power terms) and by a matched filter search (MF).
The ratio of the corresponding amplitudes is (by definition) the {\em efficiency} of the excess power statistic compared to matched filtering:
\begin{eqnarray} \label{eq:eta}
  \eta_\mathrm{EPMF}(\mathrm{FAP},\mathrm{FDP},N_\text{eff},V) &=&
  A^\mathrm{MF}_\text{min} /  A^\mathrm{EP}_\text{min}   \label{eq:eta1} \\
  \eta_\mathrm{CPEP}(\mathrm{FAP},\mathrm{FDP},V) &=&
  A^\mathrm{EP}_\text{min} /  A^\mathrm{CP}_\text{min}. \label{eq:eta3}
\end{eqnarray}

\begin{figure}
  \begin{center}
      \includegraphics[width=1.0\columnwidth]{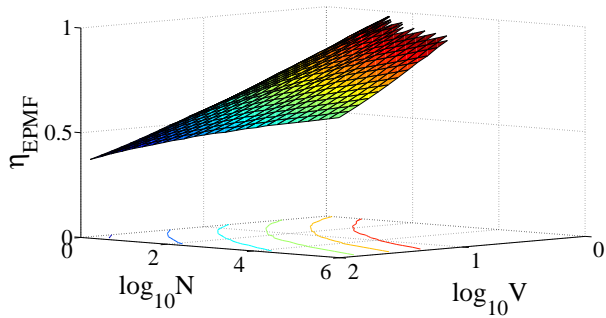} 
  \end{center}
  \caption{
    The relative detection efficiency $\eta_\mathrm{EPMF}$ of excess-power 
    compared to matched
    filtering as a function of $ft$-volume $V$ and the number of 
    effective templates $N_\text{eff}$
    (for $\mathrm{FAP} = 10^{-2}$ and $\mathrm{FDP} = 10^{-1}$).
    We assume Gaussian white noise.
    For a given value of $V$, there is maximum number for $N_\text{eff}$ 
    above which the templates will not be independent.
    Therefore, at high values of $N_\text{eff}$ and small values of $V$, 
    there is an unphysical region where $\eta>1$.
    A realistic excess-power (and hence cross-power) search might have 
    $V \sim {\cal O}(100)$, which corresponds to an 
    efficiency of $\sim50\%$ compared to matched filtering. 
    \label{fig:eta1}}  
\end{figure}

To calculate these thresholds, signal and noise distributions for CP and EP are generated using Monte Carlo simulations.
Throughout this section we assume stationary Gaussian white noise; simulated signals are characterized only by their amplitude, polarization and $ft$-volume.
Other characteristics such as frequency content, evolution with time, etc. are not relevant for this white-noise calculation.
Following~\cite{box}, we approximate the MF threshold as
\begin{equation}
  A_\text{min}^\mathrm{MF} \approx 
  A_\text{min}^\mathrm{EP}(\mathrm{FAP}/N_\text{eff}, \mathrm{FDP}, 1/2) .
\end{equation}
In Fig.~\ref{fig:eta1}, we plot $\eta_\mathrm{EPMF}$ as a function of $V$ and $N_\text{eff}$.
We see that $\eta_\mathrm{EPMF}\gtrsim50\%$ over the range of parameter space considered.

\begin{figure}
  \begin{center}
      \includegraphics[width=1.0\columnwidth]{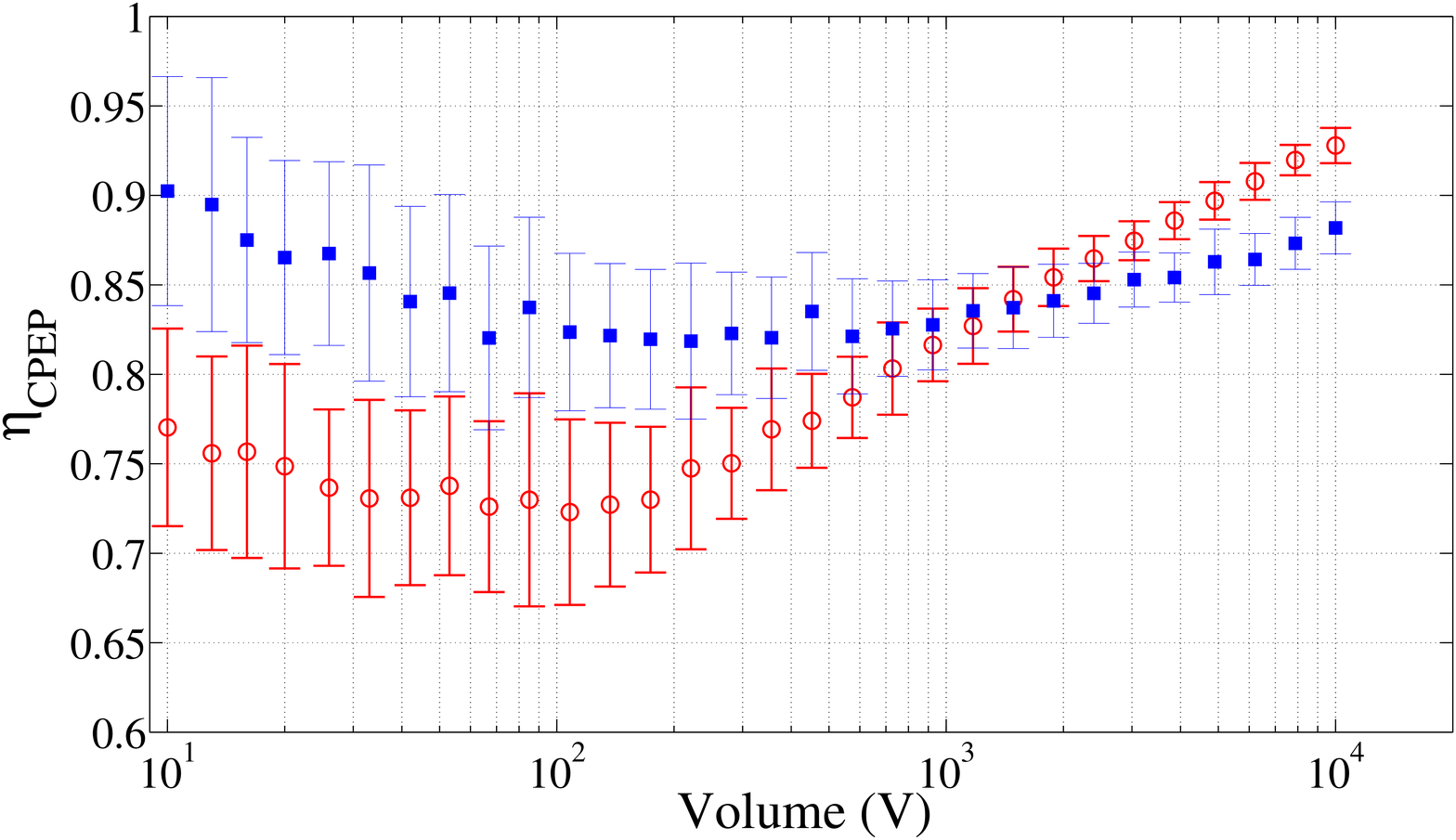} 
  \end{center}
  \caption{
    The relative detection efficiency comparing the general excess power and 
    cross-power methods for 
    $\mathrm{FAP} =10^{-2}$ and $\mathrm{FDP} = 0.1$ 
    in simulated Gaussian white noise.
    The red circles are for the H1V1 detector pair and the blue squares 
    are for the more nearly aligned H1L1 detector pair.
    The error bars represent one standard deviation using 100 trials.
    The cross-power method has a relative
    detection efficiency of at least $70\%$ with respect to the excess
    power statistic and the efficiency increases with $V$. 
\label{fig:eta3}
  } 
\end{figure}

To compare the CP method to the EP method, we calculate $\eta_\mathrm{CPEP}$ for the Hanford-Livingston and Hanford-Virgo networks averaging over an isotropically distributed population of unpolarized GW sources.
In Fig.~\ref{fig:eta3} we plot $\eta_\mathrm{CPEP}$ as a function of $ft$-volume $V$.
The CP technique is highly efficient ($\eta_\mathrm{CPEP}\sim77\%$) even at small values of $V$, but it becomes increasingly more efficient at higher $V$.

\section{Conclusions}\label{conclusions}
We reviewed a variety of compelling scenarios for long-GW transients
including protoneutron
star convection, accretion disk fragmentation/excitations, rotational
instabilities in neutron stars, $r$-modes, 
pulsar glitches and 
soft gamma repeater flares.  Many of the models we considered predict
strain amplitudes detectable in the advanced-detector era.

Next, inspired by stochastic analyses, we introduced a novel
framework, which can be used to look for GW transients on timescales
of seconds to weeks.  This framework, which is a generalization of
the GW radiometer~\cite{radiometer}, utilizes $ft$-maps of GW-strain
cross-power using two or more spatially separated interferometers
in order to look for statistically significant clustering.  A
comparison of simulated detector noise 
with time-shifted data revealed that
$ft$-cross-power-maps made with real interferometer data are
well-behaved (for at least one pixel size) suggesting that the
threshold for candidate events can be determined analytically.

We illustrated how different pattern-recognition techniques can be
used to identify GW signatures in $ft$-maps.  We demonstrated some of
these techniques using $ft$-maps generated using a GW-strain channel
cross-correlated with a LIGO microphone channel and we presented a novel
technique for the identification of environmental noise transients in
GW interferometers.
We assessed the effectiveness of our proposed statistic
compared to matched filtering and other excess-power strategies.  We
found that our strategy is highly effective when looking
for long GW transients whose precise waveforms are not known.

Many of the sources we considered here are plausible targets for the advanced detector era.
If long GW transients are, in fact, detected, we shall gain invaluable information about objects and processes, for which we currently possess only preliminary models, e.g., long gamma-ray bursts.
If, on the other hand, no long GW transients are detected, we expect that some models predicting relatively large strain amplitudes (e.g.,~\cite{vanPutten}) may be ruled out or constrained.
Third-generation detectors such as the proposed Einstein Telescope~\cite{ET} can apply our long GW transient algorithm to probe still fainter sources.

\section{Acknowledgments}
\begin{acknowledgments}
  This work was supported by NSF grants: 
  PHY-0854790, 
  PHY-0758035  
  AST-0855535, OCI-0905046, PHY-0960291 
  and
  PHY-0970074. 
  S.~G. acknowledges the support of the Max Planck
  Gesellschaft.
  P.~R. acknowledges the support of the Hungarian National Office for Research
  and Technology (NKTH) through the Polanyi program (grant KFKT-2006-01-0012).
  This paper has been assigned LIGO document number LIGO-P1000124.
\end{acknowledgments}

\begin{appendix}
\section{Derivations}\label{derivation_im}
We describe how a GW source can be characterized by its power spectrum $H(t;f)$, we construct an estimator $\hat Y(t;f)$ for $H(t;f)$ and calculate the associated variance.
We construct an estimator for the variance.
Finally, we construct the filter function for an elliptically polarized source.

\subsection{Introduction and notation}\label{GW_power}
Working in the transverse-traceless gauge we write down the general form of a GW field, which can depend on direction $\hat\Omega$, polarization state $A$ and frequency $f$ (see, e.g.,~\cite{allen-romano,radiometer,sph}):
\begin{equation}\label{app_eq:GW-wave}
  \begin{split}
    h_{ab}(t,\vec{x}) = \sum_A \int_{-\infty}^{\infty} df \int_{S^2} 
    d \hat{\Omega} \: e^A_{ab}(\hat{\Omega}) \tilde h_A(f,\hat{\Omega}) \\
    e^{2 \pi i f (t + \hat{\Omega} \cdot \vec{x} /c)} .
  \end{split}
\end{equation}
Here $\vec{x}$ and $\hat{\Omega}$ are defined in the reference frame with the
origin fixed at the center of Earth, but not rotating with the Earth.
The indices $ab$ run over a Cartesian coordinate system.
We define unit vectors
\begin{eqnarray}
  \hat{\theta} & = & \cos \theta \cos \phi \, \hat{x} + 
  \cos \theta \sin \phi \, \hat{y} - \sin \theta \, \hat{z} \\
  \hat{\phi} & = & - \sin \phi \,\hat{x} + \cos \phi \,\hat{y} \\
  \hat{\Omega} & = & \sin \theta \cos \phi \,\hat{x} + 
  \sin \theta \sin \phi \,\hat{y} + \cos \theta \,\hat{z}
\end{eqnarray}
such that $\{\hat\theta, \hat\phi, \hat\Omega\}$ form a right-handed 
coordinate system and the rotational axis of the Earth points along $\hat z $.
The two GW polarization tensors can be written as (see, e.g.,~\cite{allen-romano,radiometer,sph}):
\begin{eqnarray}
  e_{ab}^+(\hat{\Omega}) & = & \hat{\theta} \, \otimes \, \hat{\theta} - 
  \hat{\phi} \, \otimes \, \hat{\phi} \\
  e_{ab}^{\times}(\hat{\Omega}) & = & \hat{\theta} \, \otimes \, \hat{\phi} + 
  \hat{\phi} \, \otimes \, \hat{\theta} .
\end{eqnarray}

Since we are looking for GW transients, we restrict our attention to {\em point sources} for which $h_A(f,\hat\Omega)=h_A(f)\delta(\hat\Omega-\hat\Omega_0)$.
We perform the integral over $\hat\Omega$ and obtain:
\begin{equation}\label{app_eq:signal_model}
  \begin{split}
    h_{ab}(t,\vec x) = \sum_A \int_{-\infty}^\infty df \,  
    e^A_{ab}(\hat\Omega_0) \tilde h_A(f)
    \, e^{2\pi i f(t+\hat\Omega_0\cdot\vec x /c)} .
  \end{split}
\end{equation}
For simplicity, we henceforth replace $\hat\Omega_0$ with $\hat\Omega$.
It follows that the GW strain in detector $I$ is given by
\begin{equation}\label{app_eq:GW_strain}
 {h}_I(t) = \sum_A \int_{-\infty}^{\infty} df \, 
 \tilde h_A(f,\hat{\Omega}) \, e^{2 \pi i f (t+\hat{\Omega} \cdot 
   \vec{x}_I /c)} \, e^A_{ab}(\hat{\Omega}) \, d_I^{ab}(t)
\end{equation}
where $d_I^{ab}(t)$ is the detector response tensor at time t:
\begin{equation}
 d_I(t) \equiv \frac{1}{2} \left(\hat{X}_I(t) \, \otimes \, \hat{X}_I(t) - \hat{Y}_I(t) \, \otimes \, \hat{Y}_I(t)\right) .
\end{equation}
Here, the two detector arms lie along the $\hat{X}(t)$ and $\hat{Y}(t)$ axes, which are time-dependent due to the rotation of the Earth.

We now consider a finite stretch of $h_I(t)$ and take the discrete Fourier transform of Eq.\ref{app_eq:GW_strain} to obtain:
\begin{equation}\label{app_eq:h_tilde}
 \tilde{h}_I(t;f) = 
 \sum_A  \tilde h_A(t;f,\hat\Omega) \, 
 e^{2 \pi i f \hat\Omega \cdot \vec{x}_I /c} 
 \, F_I^A(t; \hat\Omega) .
\end{equation}
where we define the ``antenna factors'' (see e.g.,~\cite{allen-romano}) to be
\begin{equation}
  F_I^A(t; \hat{\Omega}) \equiv e^A_{ab}(\hat{\Omega}) \, d_I^{ab}(t) .
\end{equation}

We define the GW strain power spectrum to be
\begin{equation}\label{app_eq:spectrum}
  \langle \tilde{h}_A^*(t;f) \tilde{h}_{A'}(t;f) \rangle = 
  \frac{1}{2} H_{AA'}(t;f) ,
\end{equation}
where the factor $1/2$ comes from the fact that $H_{AA'}(t;f)$ is the one-sided power spectrum.
Here we use the discrete Fourier transform defined in Eq.~\ref{eq:DFT}; (see also Tab.~\ref{tab:DFT}).

It is convenient to characterize the source with a single spectrum that includes contributions from both $+$ and $\times$ polarizations.
We therefore define
\begin{equation}\label{app_eq:power}
  H(t;f) \equiv \text{Tr}\left[H_{AA'}(t;f)\right] ,
\end{equation}
so as to be invariant under change of polarization bases.
This definition is a generalization of the one-sided power spectrum for unpolarized sources found in~\cite{allen-romano,radiometer,sph}.
Our goal now is to derive an estimator for $H(t;f)$ in a data segment over which it is presumed to be constant.

\subsection{Derivation of $\hat Y$}\label{Y_derivation}
Let $s_I(t) = h_I(t) + n_I(t)$ be the strain time series from detector $I$,
where $h_I(t)$ is the GW strain and $n_I(t)$
is the detector noise.
Following~\cite{allen-romano,radiometer,sph}, we combine the strain time series from two spatially separated detectors, $s_I(t), s_J(t)$, to construct an estimator for GW-power $H(t;f)$ for a point source at a sky position $\hat\Omega$,
\begin{equation}
  \hat Y(t;f,\hat{\Omega}) \equiv 
  2 \,
  \text{Re}\left[ 
    \tilde{Q}_{IJ}(t;f,\hat\Omega)
    \tilde{s}_I^*(t;f) \tilde{s}_J(t;f) 
    \right]
\end{equation}
where $\tilde{Q}_{IJ}(t;f,\hat\Omega)$ is some filter function to be determined below.
We take the real part to ensure physicality of the estimator.
The expectation value of $\hat Y(t;f,\hat\Omega)$ is given by
\begin{equation}\label{app_eq:expectY}
   \langle \hat Y(t;f,\tilde{\Omega}) \rangle = 
   2 \,
   \text{Re}\left[
   \tilde{Q}_{IJ}(t;f,\hat{\Omega}) 
   \langle \tilde{h}_I^*(t;f) \, \tilde{h}_J(t;f) \rangle \, 
   \right] ,
\end{equation}
since, by assumption, there is no correlation between signal and noise and also no correlation between noise in two spatially separated detectors.

Combining Eqs.~\ref{app_eq:h_tilde},~\ref{app_eq:expectY} and~\ref{app_eq:spectrum} we get
\begin{equation}\label{app_eq:expectation}
  \begin{split}
    \langle \hat Y(t;f,\hat{\Omega}) \rangle =&
    2 \,
    \text{Re} \Big[
    \tilde{Q}_{IJ}(t;f,\hat{\Omega})
    \sum_{AA'}
    \frac{1}{2} H_{AA'}(t;f) \,
    \\
    & 
    e^{-2 \pi i f (\hat{\Omega} \cdot ({\vec{x}_{I}-\vec{x}_{J}})/c)} \,
    F_I^A(t; \hat{\Omega}) \, F_J^{A'}(t; \hat{\Omega})
    \Big]
  \end{split}
\end{equation}
In order to simplify the form of $H_{AA'}(t;f)$ we now consider {\em unpolarized} sources, for which
\begin{equation}\label{app_eq:unpolarized_model}
  H_{AA'}(t;f)=\frac{1}{2}H(t;f)\delta_{AA'} .
\end{equation}
(We consider the case of polarized sources in Subsec.~\ref{polarized}.)
For unpolarized sources,
\begin{equation}
  \begin{split}\label{app_eq:Y_is_real}
    \langle \hat Y(t;f,\hat{\Omega}) \rangle &= 
    \frac{1}{2}
    \text{Re}\Big[
      \tilde{Q}_{IJ}(t;f,\hat{\Omega}) \,
      H(t;f) \, e^{-2 \pi i f \hat{\Omega} \cdot \Delta \vec{x}_{IJ} /c} 
      \\
      & 
      \sum_A F_I^A(t; \hat{\Omega}) \, F_J^{A}(t; \hat{\Omega}) 
      \Big] ,
  \end{split}
\end{equation}
where we have defined $\Delta \vec{x}_{IJ} \equiv \vec{x}_{I} - \vec{x}_{J}$.

We desire that $\langle \hat Y \rangle = H(t;f)$, which implies:
\begin{equation}\label{app_eq:Q_unpolarized}
  \tilde Q_{IJ}(t;f,\hat\Omega) = 
  \frac{2 \, e^{2 \pi i f \hat{\Omega} \cdot \Delta \vec{x}_{IJ} /c} } 
       {\sum_A F_I^A(t; \hat{\Omega}) \, F_J^{A}(t; \hat{\Omega})} .
\end{equation}
By setting $Q_{IJ}(t;f,\hat\Omega)$ thusly, we account for the phase difference between detectors $I$ and $J$ ensuring that the bracketed quantity in Eq.~\ref{app_eq:Y_is_real} is real.
We also account for the detector pair efficiency.

Finally, we define (unpolarized) {\em pair efficiency} as
\begin{equation}\label{app_eq:efficiency}
  \epsilon_{IJ}(t;\hat\Omega) \equiv 
  \frac{1}{2} \sum_A F_I^A(t; \hat{\Omega}) 
  F_J^{A}(t; \hat{\Omega}) ,
\end{equation}
which enables us to rewrite the filter function as
\begin{equation}
  \tilde Q_{IJ}(t;f,\hat\Omega) = 
  \frac{1}{\epsilon_{IJ}(t;\hat\Omega)} e^{2 \pi i f \hat{\Omega} 
    \cdot \Delta \vec{x}_{IJ} /c} .
\end{equation}

Since $\hat Y(t;f,\hat\Omega) \propto \tilde Q(t;f,\hat\Omega)$ and $\tilde Q\propto 1/\epsilon_{IJ}(t;f,\hat\Omega)$, it follows that $\hat Y(t;f,\hat\Omega)\propto 1/\epsilon_{IJ}(t;\hat\Omega)$.
This can be understood as follows.
If we observe a modest value of strain power from a direction associated with low efficiency, we may infer (if the signal is statistically significant) that the true source power is much higher because the network only ``sees'' some fraction of the true GW power.

\subsection{Variance of the estimator}\label{variance}
We derive an expression for the variance of $\hat Y(t;f,\hat\Omega)$, $\sigma_Y(t;f,\hat\Omega)^2 \equiv \langle \hat Y(t;f,\hat\Omega)^2 \rangle - \langle \hat Y(t;f,\hat\Omega) \rangle^2$.
In searches for persistent stochastic GWs, the second term is usually omitted and the first term is simplified by assuming that signal in each pixel is small compared to the noise.
Such small signals are extracted by  averaging over a very large number of segments (see, e.g.,~\cite{allen-romano}).
Since we are dealing with transients, however, the signal may be comparable to the noise and so we can not neglect any terms in our calculation of $\sigma_Y^2$.

To begin we define a new (complex-valued) estimator that will be handy in our derivation of $\sigma_Y^2$:
\begin{equation}
  \hat W(t;f,\hat\Omega) \equiv 
  2 \,
  \tilde Q_{IJ}(t;f,\hat\Omega) \tilde s_I^\star(t;f)
  \tilde s_J(t;f) .
\end{equation}
Our GW power estimator $\hat Y(t;f,\hat\Omega)$ is simply the real part of $\hat W(t;f,\hat\Omega)$:
\begin{equation}
  \hat Y(t;f,\hat\Omega) = \frac{1}{2} \left( \hat W(t;f,\hat\Omega) + \hat W(t;f,\hat\Omega)^\star \right) .
\end{equation}
For notational compactness, we shall omit the arguments of $\hat W(t;f,\hat\Omega)$ in the remainder of this derivation.
It follows that the variance of $\hat Y(t;f,\hat\Omega)$ can be written as
\begin{equation}\label{app_eq:three_terms}
  \begin{split}
    \sigma_Y^2 = &
    \frac{1}{4} \Big[
      \left( \langle \hat W^2 \rangle - \langle \hat W \rangle^2 \right) + 
      \left( \langle \hat W^{\star2} \rangle - \langle \hat W^\star \rangle^2 \right) + 
      \\
      & 
      2\, \sigma_W^2
      \Big] ,
  \end{split}
\end{equation}
where 
\begin{equation}
  \sigma_W^2 \equiv \langle |\hat W|^2 \rangle - |\langle \hat W \rangle |^2 .
\end{equation}

Now we evaluate the three terms in Eq.~\ref{app_eq:three_terms} beginning with $\sigma_W^2$.
We obtain
\begin{equation}\label{app_eq:defn_sig_W}
  \begin{split}
    \sigma_{W}^2(t;f,\hat\Omega) &= 
    4 \,
    \Big[ \langle \tilde s_I^*(t;f) \, \tilde s_J(t;f) \, \tilde s_I(t;f) 
      \, \tilde s_J^*(t;f) \rangle  \\
      &
      - \langle \tilde s_I^*(t;f) \, \tilde s_J(t;f) \rangle \langle 
      \tilde s_I(t;f) \, \tilde s_J^*(t;f) \rangle \Big] \\
    & \left| \tilde{Q}_{IJ}(t;f,\hat{\Omega}) \right|^2 .
  \end{split}
\end{equation}
For mean-zero Gaussian random variables, we can expand the four-point correlation into a sum of products of two-point correlations.
We substitute $s=h+n$ and set signal-noise cross terms to zero along with noise-noise cross terms from different detectors.
The variance becomes
\begin{equation}\label{app_eq:sigma}
  \begin{split}
    \sigma_{W}^2(t;f,\hat\Omega) = &
    4 \, \Big[ 
      \langle \tilde{h}_I^*(t;f) \tilde{h}_I(t;f) \rangle \: 
      \langle \tilde{h}_J(t;f) \tilde{h}_J^*(t;f) \rangle + \\
      & \langle \tilde{h}_I^*(t;f) \tilde{h}_I(t;f) \rangle \: 
      \langle \tilde{n}_J(t;f) \tilde{n}_J^*(t;f) \rangle + \\ 
      &
      \langle \tilde{h}_J(t;f) \tilde{h}_J^*(t;f) \rangle 
      \: 
      \langle \tilde{n}_I^*(t;f) \tilde{n}_I(t;f) \rangle
      + \\
      & \langle \tilde{n}_I^*(t;f) \tilde{n}_I(t;f) \rangle \: 
      \langle \tilde{n}_J(t;f) \tilde{n}_J^*(t;f) \rangle \Big] \\
    & \left| \tilde{Q}_{IJ}(t;f,\hat{\Omega}) \right|^2,
  \end{split}
\end{equation}

Evaluating the four terms in Eq.~\ref{app_eq:sigma}, we obtain
\begin{equation}\label{app_eq:var_W}
  \begin{split}
    \sigma_{W}^2(t;f,\hat\Omega) = 
    \bigg[
      \epsilon_{II}(t;\hat\Omega) \epsilon_{JJ}(t;\hat\Omega) H(t;f)^2 + 
      \\
      H(t;f) \left(
      \epsilon_{II}(t;\hat\Omega) N_J(t;f) \, +
      \epsilon_{JJ}(t;\hat\Omega) N_I(t;f)  
      \right)
      \\
      + \, N_I(t;f,)N_J(t;f)
      \bigg]
    \, \left|\tilde Q_{IJ}(t;f,\hat\Omega) \right|^2 ,
  \end{split}
\end{equation}
where $\epsilon$ is defined in Eq.~\ref{app_eq:efficiency} and where $N_I(t;f)$ is the one-sided noise-power spectra:                         
\begin{equation}
  N_I(t;f) \equiv 2 \, \left| \tilde n_I(t;f) \right|^2 .
\end{equation}

Using the same line of reasoning, we calculate the remaining terms in Eq.~\ref{app_eq:three_terms}:
\begin{equation}\label{app_eq:correction_terms}
  \begin{split}
    \langle \hat W^2 \rangle - \langle \hat W \rangle^2  = 
    \langle \hat W^{\star2} \rangle - \langle \hat W^\star \rangle^2  = 
    H(t;f)^2 .
  \end{split}
\end{equation}
Combining Eqs.~\ref{app_eq:three_terms} and~\ref{app_eq:correction_terms}, we conclude that
\begin{equation}\label{app_eq:sig_Y_final}
  \sigma_Y^2 = \frac{1}{2} \left[ \sigma_W^2 + H(t;f)^2 \right] .
\end{equation}
The factor of $1/2$ comes about from the fact that $\hat Y(t;f,\hat\Omega)$ is real whereas $\hat W(t;f,\hat\Omega)$ is complex.
We note that in the small-signal limit $H(f)\rightarrow0$ and the variance reduces to the canonical stochastic result~\cite{allen-romano}:
\begin{equation}
  \sigma_Y^2 \rightarrow \frac{1}{2} \left(
  N_I(t;f) N_J(t;f) \left| \tilde Q_{IJ}(t;f,\hat\Omega) \right|^2
  \right) ,
\end{equation}

\subsection{Expectation value of $\hat{\sigma}_Y^2$}\label{sigma_hat}
Our estimator for the variance of $\hat Y$ is given by
\begin{equation}\label{app_eq:pixelvar}
  \begin{split}
    \hat\sigma_Y^2(t;f,\hat\Omega) = \frac{1}{2}
    \left|\tilde Q_{IJ}(t;f,\hat\Omega)\right|^2
    P_I^\text{adj}(f)P_J^\text{adj}(f) ,
  \end{split}
\end{equation}
where $P_I$ is the average auto-power in neighboring pixels:
\begin{equation}\label{app_eq:PSD_def}
  P_I^\text{adj}(f) \equiv 2 \,
  \overline{ \left| \tilde s_I(f) \right|^2 } .
\end{equation}
The overline denotes an average over neighboring pixels.
By averaging over neighboring pixels, we assume that the detector noise in any given pixel can be characterized by looking at its neighbors.
This assumption is discussed below.

Now we calculate the expectation value of our estimator for variance $\hat\sigma_Y^2$ given in Eq.~\ref{app_eq:pixelvar} in order to compare it to the theoretical variance given in Eqs.~\ref{app_eq:sig_Y_final} and~\ref{app_eq:var_W}.
Eqs.~\ref{eq:pixelvar} and~\ref{eq:PSD_def} together imply
\begin{equation}
  \begin{split}
    \langle \hat\sigma_Y^2(t;f,\hat\Omega) \rangle 
    = & 2 \,
    \left|\tilde Q_{IJ}(t;f,\hat\Omega)\right|^2 \\
    &
    \langle s_I^{*\text{adj}}(f)s_I^{\text{adj}}(f)s_J^{*\text{adj}}(f)
    s_J^{\text{adj}}(f) \rangle.
  \end{split}
  \label{eq:start_exp_sigma}
\end{equation}
Using Equation~\ref{app_eq:defn_sig_W} to write the expectation value of $\hat{\sigma}_Y^2$ in terms of the theoretical value of $\sigma_W^2$, we find
\begin{equation}
  \begin{split}
    \langle \hat\sigma_Y^2(t;f,\hat\Omega)\rangle
    = &\frac{1}{2} 
    \left[
      \sigma_W^2(t;f,\hat{\Omega}) +
      4 \,
      \left|\tilde{Q}_{IJ}(t;f,\hat{\Omega})\right|^2  \right.\\
      & \left .
      \langle\tilde{s}^\text{adj*}_I(t;f)\tilde{s}^\text{adj}_J(t;f)\rangle
      \langle\tilde{s}^\text{adj}_I(t;f)\tilde{s}^\text{adj*}_J(t;f)
      \rangle\right]\\
    =&\frac{1}{2}\left[\sigma_W^2(t;f,\hat{\Omega})
      +\left|\langle\hat{W}\rangle\right|^2\right] \\
        =&\frac{1}{2}\left[\sigma_W^2(t;f,\hat{\Omega})
      +H(t;f)^2\right]
\end{split}
\label{eq:exp_sig_Y_int_step}
\end{equation}
Since this is the theoretical variance from \ref{app_eq:sig_Y_final}, we
conclude that $\left\langle \hat{\sigma}_Y^2\right\rangle=\sigma_Y^2$.
Thus, Eq.~\ref{eq:pixelvar} provides an unbiased estimator for $\sigma_Y^2$.
Here we have assumed that the noise and signal are comparable in neighboring segments.
This assumption can fail for rapidly changing, high-$\text{SNR}$ signals and also for highly non-stationary noise, and so additional work may be required to estimate $\sigma$ in these situations.

\subsection{Elliptically polarized sources}\label{polarized}
A variety of long-transient GW sources are expected to be elliptically polarized (e.g., long GRBs~\cite{vanPutten,piro} and pulsar glitches~\cite{vanEysden}).
Elliptically polarized sources are parameterized by two angles.
The inclination angle $\iota$ is the angle between the rotational axis of the source and the observer's line of sight and the polarization angle $\psi$ describes the orientation of the rotational axis in the plane perpendicular to the line of sight (see, e.g.,~\cite{prix}).

Following~\cite{prix-whelan}, we characterize an elliptically polarized source with the so-called canonical amplitudes:
\begin{equation}
  {\cal A}_\mu \equiv \left(
  \begin{array}{c}
    A_+ \cos 2\psi \\
    A_+ \sin 2\psi \\
    -A_\times \sin 2\psi \\
    A_\times \cos 2\psi ,
  \end{array}
  \right)
\end{equation}
where
\begin{eqnarray}\label{app_eq:A+Ax}
  A_+ & \equiv & \left( h_0/2\right) \left(1+\cos^2 \iota \right) \\
  A_\times & \equiv & h_0 \cos\iota
\end{eqnarray}
We have set the initial phase $\phi_0=0$ for the sake of simplicity.
(Ultimately, we are concerned with the average cross-power over many cycles and so the initial phase is unimportant.)
Here $h_0$ is the strain amplitude.

Next, we define the tensor~\cite{prix-whelan}
\begin{equation}
  h_{ab}^\mu(t) \equiv \left(
  \begin{array}{c}
    e^+_{ab} \cos\left[\phi(t)\right] \\
    e^\times_{ab} \cos\left[\phi(t)\right] \\
    e^+_{ab} \sin\left[\phi(t)\right] \\
    e^\times_{ab} \sin\left[\phi(t)\right]
  \end{array}
  \right) ,
\end{equation}
where $\phi(t)$ describes the phase evolution of the signal~\footnote{The reader is advised to use caution when comparing our formulas with those in~\cite{prix-whelan}.
Following~\cite{allen-romano,radiometer,sph}, we use $e^A_{ab}$ to denote the ``detector polarization basis'' defined by the ecliptic plane.
In~\cite{prix-whelan}, however, this quantity is denoted $\epsilon_{ab}^A$ and $e_{ab}^A$ is the basis in which the plus polarization is maximal.
}.
Now we write the GW equation as~\cite{prix-whelan}
\begin{equation}\label{app_eq:polar_gw}
  \begin{split}
    h_{ab}(t) & \equiv \sum_\mu {\cal A}_\mu h^\mu_{ab}(t) = \\
    & \left( A_+ \cos(2\psi) \cos\left[\phi(t)\right]
    -A_\times \sin(2\psi) \sin\left[\phi(t)\right] \right) e_{ab}^+ + \\
    & \left( A_+ \sin(2\psi) \cos\left[\phi(t)\right] +
    A_\times \cos(2\psi) \sin\left[\phi(t)\right] \right) e_{ab}^\times .
  \end{split}
\end{equation}

We Fourier transform a finite stretch of GW signal to obtain the coefficients $\tilde h_A(f)$ (see Eq.~\ref{app_eq:GW-wave}):
\begin{eqnarray}\label{app_eq:polar_fourier_coeff}
  \tilde h_+(t;f) & = & \frac{
    \left[ A_+ \cos(2\psi) + i A_\times \sin(2\psi) \right] \delta_{ff_0} 
  }{
    2N_s
  }
  \\
  \tilde h_\times(t;f) & = & \frac{
    \left[ A_+ \sin(2\psi) - i A_\times \cos(2\psi) \right]
    \delta_{ff_0} 
  }{
    2N_s
  }
    .
\end{eqnarray}
We have ignored negative frequencies, which will play no part in our subsequent calculation of one-sided GW power.
Also, we have expanded $\phi(t)$ in a Taylor series,
\begin{equation}
  \phi(t) = \phi_0 + 2\pi \left(f_0 t + \frac{1}{2} \dot f t^2 \right) + 
      {\cal O}(t^3) .
\end{equation}
We assume that $\dot f$ is approximately zero over the segment duration, $T$.
In principle, this formulation could be extended to broadband sources, but we expect elliptically polarized sources (associated with spinning objects and a binary objects) to be narrowband.

Combining Eqs.~\ref{app_eq:spectrum} and~\ref{app_eq:polar_fourier_coeff}, we calculate the one-sided power spectrum:
\begin{widetext}
  \begin{equation}\label{app_eq:H_polar_matrix}
    \begin{split}
      H_{AA'}(t;f) = & \frac{\delta_{ff_0}}{2 N^2_s}
      \left[
	\left(
	\begin{array}{cc}
	  A_+^2 \cos(2\psi)^2 + A_\times^2 \sin(2\psi)^2 &
	  (A_+^2-A_\times^2) \cos(2\psi) \sin(2\psi) \\
	  (A_+^2-A_\times^2) \cos(2\psi) \sin(2\psi) &
	  A_+^2 \sin(2\psi)^2 + A_\times^2 \cos(2\psi)^2
	\end{array}
	\right) 
	+ i
	\left(
	\begin{array}{cc}
	  0 & -A_+ A_\times \\
	  A_+ A_\times & 0
	\end{array}
	\right)
	\right]
    \end{split}
  \end{equation}
\end{widetext}
The imaginary off-diagonal terms corresponds to the phase delay between $+$ and $\times$ polarization states.

Combining Eqs.~\ref{app_eq:power} and~\ref{app_eq:H_polar_matrix}, we obtain
\begin{equation}\label{app_eq:H_polar}
  H(t;f) = \frac{1}{2 N^2_s} \left(A_+^2 + A_\times^2 \right) \, 
  \delta_{ff_0}.
\end{equation}
Our goal, once again, is to find an estimator for $H(t;f)$.
We assume that the estimator for $H(t;f)$ can be constructed from the cross power spectrum of two GW-strain channels multiplied by an appropriate filter function, $\tilde Q_{IJ}(t;f,\hat\Omega,\iota,\psi)$:
\begin{equation}\label{app_eq:polarized_estimator}
  \begin{split}
    \hat Y(t;f,\hat\Omega,\iota,\psi) = 
    2 \,
    \text{Re}\left[
      \tilde s_I^\star(t;f) \tilde s_J(t;f) \,
      \tilde Q_{IJ}(t;f,\hat\Omega,\iota,\psi) \right] .
  \end{split}
\end{equation}

Plugging Eq.~\ref{app_eq:H_polar_matrix} into Eq.~\ref{app_eq:expectation}, we obtain
\begin{widetext}
  \begin{equation}\label{app_eq:expect_Y_polar}
    \begin{split}
      \left\langle \hat Y(t;f,\hat\Omega,\iota,\psi) \right\rangle & =
      2 \,
      \text{Re} 
      \bigg[
	\frac{1}{2}
	\frac{ \delta_{ff_0} }{2N^2_s}
	\Big( 	
	F_I^+ F_J^+ 
	\left[ A_+^2 \cos(2\psi)^2 + A_\times^2 \sin(2\psi)^2 \right]
	\\
	&
	+
	F_I^+ F_J^\times 	
	\left[ (A_+^2-A_\times^2) \cos(2\psi) \sin(2\psi) -iA_+A_\times \right]
	+
	F_I^\times F_J^+ 
	\left[ (A_+^2-A_\times^2) \cos(2\psi) \sin(2\psi) +iA_+A_\times \right]
	\\
	&
	+
	F_I^\times F_J^\times
	\left[ A_+^2 \sin(2\psi)^2 + A_\times^2 \cos(2\psi)^2 \right]
	\Big) 
	\, 
	e^{-2\pi i f(\hat\Omega\cdot\Delta \vec x_{IJ}/c )} \,
	\tilde Q_{IJ}(t;f,\hat\Omega,\iota,\psi)
	\bigg] .
    \end{split}
  \end{equation}
\end{widetext}

We write the polarized filter function as:
\begin{equation}\label{app_eq:Q_polar}
  \begin{split}
    \tilde Q_{IJ}(t;f,\hat\Omega,\iota,\psi) = 
    e^{2\pi i f \hat\Omega\cdot\Delta\vec x_{IJ}/c+i\eta} / \epsilon(\hat\Omega,\iota,\psi) ,
  \end{split}
\end{equation}
where $\eta$ is an angle arising from from the phase delay between $+$ and $\times$ polarizations.
By requiring that $\langle \hat Y(t;f,\hat\Omega,\iota,\psi) \rangle=H(t;f)$, it follows that
\begin{equation}
  \begin{split}
    \epsilon(\hat\Omega,\iota,\psi) & =
    \Big|
    F_I^+ F_J^+ [a_+^2 \cos(2\psi)^2 + a_\times^2 \sin(2\psi)^2] + \\
    & (F_I^+ F_J^\times + F_I^\times F_J^+) 
    (a_+^2 - a_\times^2) \cos(2\psi) \sin(2\psi) + \\
    & F_I^\times F_J^\times (a_+^2 \sin(2\psi)^2 + 
    a_\times^2 \cos(2\psi)^2) \\
    & 
    + i a_+a_\times(F_I^\times F_J^+ - F_I^+ F_J^\times)
    \Big| \\
    & /  (a_+^2 + a_\times^2) ,
  \end{split}
\end{equation}
and
\begin{equation}
  \begin{split}
    \eta(\hat\Omega,\iota,\psi) & =
    -\text{phase}\bigg(\Big[
    F_I^+ F_J^+ [a_+^2 \cos(2\psi)^2 + a_\times^2 \sin(2\psi)^2] + \\
    & (F_I^+ F_J^\times + F_I^\times F_J^+) 
    (a_+^2 - a_\times^2) \cos(2\psi) \sin(2\psi) + \\
    & F_I^\times F_J^\times (a_+^2 \sin(2\psi)^2 + 
    a_\times^2 \cos(2\psi)^2) \\
    & 
    + i a_+a_\times(F_I^\times F_J^+ - F_I^+ F_J^\times)
    \Big]
    \\
    & /  (a_+^2 + a_\times^2) 
    \bigg),
  \end{split}
\end{equation}
where we have defined
\begin{eqnarray}
  a_+ & \equiv & A_+ / h_0 \\
  a_\times & \equiv & A_\times / h_0 .
\end{eqnarray}
As a sanity check, we note that an elliptically polarized source with pure $+$-polarization yields a sensible efficiency 
\begin{equation}
  \epsilon(\hat\Omega,\iota=90^\circ,\psi=0) =
  F_I^+ F_J^+ .
\end{equation}

\section{Data processing details}
\subsection{Overview}\label{gauss_items}
In this we describe the detailed procedure for the creation of the $\unit[52]{s}\times\unit[0.25]{Hz}$ coarse-grained data in Fig.~\ref{fig:snr_pdf}.
\begin{itemize}
\item Since $n(t)$ is drawn from a zero-mean, unit-variance normal distribution, the frequency domain data will be zero-mean with a standard deviation scaled by $\sqrt{N/2}$ for the real and imaginary parts separately where $N$ is the number of sampling points.
\item To calculate the cross-spectral density, $\text{CSD}$, data are first Hann windowed and zero padded in the time domain.  Discrete Fourier transforms are multiplied to form cross-power, then summed (with a window that tapers at the ends) to yield a $\text{CSD}$ with $\unit[0.25]{Hz}$ resolution.
\item To calculate each $P_I(f)$, data are broken into $\unit[4]{s}$ blocks
with 50\% overlap.  Each block is Hann windowed, and the resulting power samples from all blocks are summed so that $P_I(f)$ also has $\unit[0.25]{Hz}$ resolution.
\item Hann windowing reduces the overall power by a factor of $3/8$.  To compensate both the $\text{CSD}$ and $P_I(f)$ are corrected for this factor.
\item Without windowing or overlap each power estimate from each data block will be $\chi^2$-distributed with 2 degrees of freedom and scaled by $N/2$, and so the average of $M$ blocks are $\chi^2$-distributed with $2M$ degrees of freedom and scaled by $N/2M$.
\item Overlap changes the effective number of degrees of freedom in a way that can be calculated numerically (see App.~\ref{gauss_app}).
\item Without windowing, each cross-power product will have zero-mean and a standard deviation scaled by $N/\sqrt{2}$. Summing $M$ values gives each $\text{CSD}$ estimate a standard deviation scaled by $N\sqrt{M/2}$.
\item These scaling behaviors indicate how we could expect $\text{SNR}$ to behave, except for the non-trivial effects of windowing, zero-padding and weighted averaging.  These somewhat subtle effects are evaluated explicitly in App.~\ref{gauss_app}.
\end{itemize}

\subsection{Calculation of the normalization factor}\label{gauss_app}
Here we derive the stretch factor applied to the signal to noise ratio ($\text{SNR}$) histogram in Fig.~\ref{fig:snr_pdf} generated using simulated Gaussian noise.
Let $\{x_{n}\}$ and $\{y_{m}\}$ be real, discrete, time sequences corresponding to two independent data streams with each $x_{n}$ and $y_{m}$ sampled from a (0,1)-Gaussian distribution. The time-domain $\{x_{n}\}$ and $\{y_{m}\}$ can be transformed into frequency-domain sequences of N complex numbers $\{X_{p}\}$ and $\{Y_{q}\}$ using the discrete Fourier transform (DFT), defined here as:
\begin{equation}
  \begin{split}
    X_{p}& = \Delta t \sum_{n=0}^{N-1} \exp\left({\frac{-2 \pi i pn}{N}}\right) x_{n}  \,\,\,\,\,\ \mbox{and} \\
    Y_{q}& = \Delta t \sum_{m=0}^{N-1} \exp\left({\frac{-2 \pi i qm}{N}}\right) y_{m}  
  \end{split}
  \label{eq:Xnozpad}
\end{equation}
The inverse DFT can be used to express the original discrete-time signal as
\begin{equation}
  \begin{split}
    x_{n} & = \Delta f \sum_{p=0}^{N-1} \exp\left({\frac{2 \pi i pn}{N}}\right) X_{p}   \,\,\,\,\,\  \mbox{and}  \\ 
    y_{m} & = \Delta f \sum_{q=0}^{N-1} \exp\left({\frac{2 \pi i qm}{N}}\right) Y_{q} 
  \end{split}
  \label{eq:xnozpad}
\end{equation}

To see the effects of zero padding, we define the real, 2N-periodic discrete-time signal $\tilde{x}\equiv \{x_n , 0_n \}$ where $0_n $ is a sequence of N-many zeros.
Then 
\begin{equation}
  \begin{split}
    \tilde{X}_{p} & = \Delta t \sum_{n=0}^{2N-1} \exp\left({\frac{-2 \pi i pn}{2N}}\right) \tilde{x}_n   \\
    & = \Delta t \sum_{n=0}^{N-1} \exp\left({\frac{-2 \pi i pn}{2N}}\right) x_{n} 
  \end{split}
\end{equation}
\mbox{where} 
\begin{equation}
  \tilde{x}_n= \begin{cases}x_n = \frac{\Delta f}{2} \sum_{p=0}^{2N-1} \exp\left({\frac{2 \pi i pn}{2N}}\right) \tilde{X}_p      
    \\\\\ 0_n = \frac{\Delta f}{2} \sum_{p=0}^{2N-1} (-1)^p \exp\left({\frac{2 \pi i pn}{2N}}\right) \tilde{X}_p 
  \end{cases}
  \label{eq:xtilde}
\end{equation}
\mbox{with} \,\,\,\,\ $0\leq n \leq N-1$
\\

From Eq.~\ref{eq:xtilde} we conclude that the $X_p \ \mbox{with} \ 0\leq p \leq 2N-1$ are not linearly independent.
However, the $X_{2p}$ $(0\leq p \leq N-1)$ and $X_{2p+1}$ $(0\leq p \leq N-1)$ are separately linearly independent.
In the calculations below, we use only the even terms to express both the even and odd $X_k$ terms. Therefore, taking into account the zero-padding, we can write 
\\
\begin{equation}
  \begin{split}
    x_{n} & = \Delta f \sum_{p=0}^{N-1} \exp\left({\frac{2 \pi i (2p)n}{2N}}\right) \tilde{X}_{2p}  \\
    & = \Delta f \sum_{p=0}^{N-1} \exp\left({\frac{2 \pi i pn}{N}}\right) \tilde{X}_{2p} 
  \end{split}
  \label{eq:xzpad}
\end{equation}
\\
By comparing $x_n$ in Eqs.~\ref{eq:xnozpad}~\ref{eq:xzpad} we see that
\begin{equation}
  \tilde{X}_{2p} = X_p
  \label{eq:X2ptilde}
\end{equation}
In addition to the zero padding, the windowing effect can be expressed by defining $\tilde{u} \equiv \{u_n , 0_n \} \equiv \{x_nf(n) , 0_n \}$ with the discrete Fourier transform given by
\\
\begin{equation}
  \begin{split}
    U_{k} & = \Delta t \sum_{n=0}^{2N-1} \exp\left({\frac{-2 \pi i kn}{2N}}\right) \tilde{u}_n   \\
    & = \Delta t \sum_{n=0}^{N-1} \exp\left({\frac{-2 \pi i kn}{2N}}\right) u_n   \\
    & = \Delta t \sum_{n=0}^{N-1} \exp\left({\frac{-2 \pi i kn}{2N}}\right) f(n) x_n  
  \end{split}
  \label{eq:Uzpadwind}
\end{equation}
\\
where $x_n$ is given by Eq.~\ref{eq:xzpad} and $f(n)$ is the Hann window defined by 
\\
\begin{equation}
  \begin{split}
    f(n) & \equiv \frac{1}{2} \left[1-\cos\left(\frac{2 \pi n}{N} \right) \right] \\
    & = \sum_{a=-1}^{1} \frac{1}{2} \left(\frac{-1}{2} \right)^{|a|} \exp\left({\frac{2 \pi i na}{N}}\right)\,\,\,\,\,\ 
  \end{split}
  \label{eq:Hannwind}
\end{equation}
$\mbox{with} \,\,\,\,\,\ 0 \leq n \leq N-1$
\\

After some algebra Eq.~\ref{eq:xzpad}, Eq.~\ref{eq:X2ptilde}, Eq.~\ref{eq:Uzpadwind} and Eq.~\ref{eq:Hannwind} give
\\
\begin{equation}
  \begin{split}
U_k = \frac{\Delta f \Delta t}{2} & \sum_{n=0}^{N-1}\sum_{p=0}^{N-1} \sum_{a=-1}^{1} X_p \\
& \exp\left({\frac{-2 \pi i n(k/2-p-a)}{N}}\right)\left( - \frac{1}{2}\right)^{|a|}
  \end{split}
  \label{eq:Uafteralg}
\end{equation}
\\
Similarly, we can express the second data stream by $\tilde{v} \equiv \{v_m , 0_m \} \equiv \{y_mf(m) , 0_m \} $ with $0 \leq m \leq N-1$. In analogy with \ref{eq:Uafteralg} the discrete Fourier transform is given by
\begin{equation}
  \begin{split}
    V_k = \frac{\Delta f \Delta t}{2} & \sum_{m=0}^{N-1} \sum_{q=0}^{N-1} \sum_{b=-1}^{1} Y_q \\
    & \exp\left({\frac{-2 \pi i m(k/2-q-b)}{N}}\right)\left( - \frac{1}{2}\right)^{|b|} 
  \end{split}
\end{equation}
\\
The sequences $\{\tilde{u}_n\}$ and $\{\tilde{v}_n\}$ are the real, 2N-periodic, discrete-time signals that are used in the code for the calculation of the contribution of the ($\text{CSD}$) to the ($\text{SNR}$).

From the definition
\begin{equation}
  \begin{split}
    SNR(t;f)= &
    \text{Re} \Big| \frac{\text{CSD}_{xy}(t;f)}{\sqrt{\text{PSD}_{x}(t;f)\text{PSD}_{y}(t;f)}} 
    \\
    & 
    \text{phase}( Q(t;f,\hat\Omega)) \Big| ,
    \label{eq:SNR}
  \end{split}
\end{equation}
\\
where $Q(t;f,\hat\Omega)$ is the filter function, we observe that in order to explain the $\text{SNR}$ distribution obtained from the processed data we need to understand the statistical consequences of the code that calculates the power spectral density $\text{PSD}_{x}$ (or $\text{PSD}_{y}$)  and the $\text{CSD}_{xy}$.\\

Since the $\text{CSD}$ is the Fourier transform of the cross-covariance function, in order to characterize the numerator of $\text{SNR}$ we need to study terms of the form $X_{k}^*Y_{k}$. Without windowing and without zero-padding the $X_{k}^*Y_{k}$ term depends only on the frequency $k$ as the following result shows (obtained using Eq.~\ref{eq:Xnozpad} and Eq.~\ref{eq:xnozpad}).
\\
\begin{equation}
  \begin{split}
    X_{k}^*Y_{k}= & \sum_{n=0}^{N-1}\sum_{m=0}^{N-1}\sum_{p=0}^{N-1}\sum_{q=0}^{N-1}X_{p}^*Y_{q} \exp\left({\frac{2 \pi i n (k-p)}{N}}\right) \\
    & \exp\left({\frac{-2 \pi i m (k-q)}{N}}\right) \\
    & =  \sum_{p=0}^{N-1} \sum_{q=0}^{N-1} \delta_{pk} \delta_{qk} X_{p}^* Y_{q}
  \end{split}
  \label{eq:Radonighte}
\end{equation}
\\
However, when we take into account the effects of windowing and zero-padding, the $U^*_k V_k$ term, contrary to the $X_{k}^*Y_{k}$ term, has contributions from a wider part of the frequency domain according to the following result
\\
\begin{equation}
  \begin{split}
    U_{k} & ^*V_{k}= \frac{1}{4} \sum_{n=0}^{N-1}\sum_{m=0}^{N-1}\sum_{p=0}^{N-1}\sum_{q=0}^{N-1}\sum_{a=-1}^{1}\sum_{b=-1}^{1}X_{p}^*Y_{q} \left(-\frac{1}{2}\right)^{|a|+|b|} \\
    &  \exp\left({\frac{2 \pi i n (k/2-p-a)}{N}}\right) \exp\left({\frac{-2 \pi i m (k/2-q-b)}{N}}\right)
    \label{eq:RadonWeigholve}
  \end{split}
\end{equation}
\\

To imitate the way the code is calculating the $\text{CSD}$, we define
\begin{equation}
  Z_{j}\equiv\sum_{k=-13}^{+13}w(k)U_{j+k}^*V_{j+k}
\end{equation}
where
\begin{equation}
  w(k)\equiv(\frac{1}{2},\underbrace{1,1,...,1,}_{25-\text{times}}  \frac{1}{2})
\end{equation}
\\
It is easy to see that the expectation value of $Z_j$ is zero, since the $X_q$ and $Y_q$ are statistically independent with mean zero.  To evaluate the standard deviation of $Z_j/\sqrt{26}$, where the 26 is a normalization constant due to averaging, we then need to evaluate $ \langle\sqrt{Z_{j}^*Z_{j}/26}\rangle$ (where there is no sum over $j$).
After some calculations, the final expression for $\langle Z_{j}^*Z_{j}\rangle$ becomes

\begin{equation}
  \begin{split}
    \langle Z_{j}^* Z_{j}\rangle = \sum_{k,k',\alpha,\alpha',\beta,\beta'} & w(k) w(k')\Phi^*(\alpha,\alpha')\Phi(\beta,\beta') \\
    & \left(\frac{1}{4}\right)^2 \left(-\frac{1}{2}\right)^{(|a|+|b|+|a'|+|b'|)}
  \end{split}
\end{equation}
where
\begin{equation}
  \begin{split}
    \Phi(\alpha,\alpha') \equiv & \frac{\sin\left[(\pi/2)(k-k'+2\alpha -2\alpha')\right]}{\sin\left[(\pi/2N)(k-k'+2\alpha-2\alpha')\right]} \\
    & \exp\left[-i\pi(\alpha-\alpha')\left(1-\frac{1}{N}\right)\right]
  \end{split}
\end{equation}
\\

Using $N=212,992$, corresponding to $\unit[52]{s}$ of data sampled at $\unit[4,096]{Hz}$, the numerical result turned out to be 
\begin{equation}
  \sqrt{\frac{1}{26}\langle Z_{j}^*Z_{j}\rangle}=0.7208 .
\end{equation}
Multiplying this by the normalization factor of $8/3$ due to Hann window we get 1.922---the normalization factor applied to MC in Fig~\ref{fig:snr_pdf}.

\end{appendix}

\bibliography{stamp}

\end{document}